%% file: main.tex
\documentclass[conference]{IEEEtran}
\IEEEoverridecommandlockouts
\usepackage{cite}
\usepackage{amsmath,amssymb,amsfonts}
\usepackage{braket}
\usepackage{caption}
\usepackage[ascii]{inputenc}
\usepackage{pgfplots}
\DeclareUnicodeCharacter{2212}{−}
\usepgfplotslibrary{groupplots,dateplot}
\usetikzlibrary{patterns,shapes.arrows}
\pgfplotsset{compat=newest}
\usepackage{subcaption}
\usepackage{balance}
\usepackage{ifthen}
\usepackage{algorithmic}
\usepackage{algorithm2e}
\usepackage{float}
\usepackage{mathtools}
\usepackage{hyperref}
\usepackage{cleveref}
\usepackage{graphicx}
\usepackage{fancyhdr}

\usepackage{fancyhdr}

\fancypagestyle{specialfooter}{%
	\fancyhf{}
	
	\fancyfoot[R]{ \noindent\fbox{%
			\parbox{\textwidth}{%
				{\footnotesize \copyright 2024 IEEE. Personal use of this material is permitted. Permission from IEEE must be obtained for all other uses, in any current or future media, including reprinting/republishing this material for advertising or promotional purposes, creating new collective works, for resale or redistribution to servers or lists, or reuse of any copyrighted component of this work in other works.}
			}
	}}
}

\def\BibTeX{{\rm B\kern-.05em{\sc i\kern-.025em b}\kern-.08em
		T\kern-.1667em\lower.7ex\hbox{E}\kern-.125emX}}
\usepackage{textcomp}
\usepackage{xcolor}
\def\BibTeX{{\rm B\kern-.05em{\sc i\kern-.025em b}\kern-.08em
		T\kern-.1667em\lower.7ex\hbox{E}\kern-.125emX}}

\begin{document}
	\pagestyle{plain}
	
	\renewcommand{\vec}[1]{\boldsymbol{\mathbf{#1}}}
	\title{Efficient Encodings of the Travelling Salesperson Problem for Variational Quantum Algorithms 
	}
	
	\author{
		\IEEEauthorblockN{Manuel Schnaus\IEEEauthorrefmark{1}\IEEEauthorrefmark{2}, Lilly Palackal\IEEEauthorrefmark{1}\IEEEauthorrefmark{2}, Benedikt Poggel\IEEEauthorrefmark{3}, Xiomara Runge\IEEEauthorrefmark{3},\\
			Hans Ehm\IEEEauthorrefmark{1}, Jeanette Miriam Lorenz\IEEEauthorrefmark{3}, Christian B. Mendl\IEEEauthorrefmark{2}}
		\IEEEauthorblockA{\IEEEauthorrefmark{1}Infineon Technologies AG, Munich, Germany}
		\IEEEauthorblockA{\IEEEauthorrefmark{2}Technical University of Munich, Germany}
		\IEEEauthorblockA{\IEEEauthorrefmark{3}Fraunhofer Institute for Cognitive Systems IKS}
		\{manuel.schnaus, lilly.palackal, hans.ehm\}@infineon.com \\
		\{xiomara.runge, benedikt.poggel, jeanette.miriam.lorenz\}@iks.fraunhofer.de \\
		christian.mendl@tum.de
	}
	
	\maketitle
	\thispagestyle{specialfooter}
	\begin{abstract}
		Routing problems are a common optimization problem in industrial applications, which occur on a large scale in supply chain planning. Due to classical limitations for solving NP-hard problems, quantum computing hopes to improve upon speed or solution quality. Several suggestions have been made for encodings of routing problems to solve them with variational quantum algorithms. However, for an end user it is hard to decide a priori which encoding will give the best solutions according to their needs. In this work, we investigate different encodings for the Travelling Salesperson Problem. We compare their scaling and performance when using the Quantum Approximate Optimization Algorithm and the Variational Quantum Eigensolver and provide a clear guide for users when to choose which encoding. For small instances, we find evidence that the permutation encoding can yield good results since it does not suffer from feasibility issues.
	\end{abstract}
	
	
	\section{Introduction}
	Solving optimization problems in complex supply chains is a big challenge for many industries. To guarantee resilient planning, various computationally intensive optimization problems have to be solved at large scale on a daily basis. This includes routing problems such as the Travelling Salesperson Problem (TSP) as well as more complex variants like the Capacitated Vehicle Routing Problem (CVRP). Given the vast potential of quantum computing for solving difficult problems, it becomes vital to investigate the applicability of quantum methods to exactly these problems, which are computationally costly for classical computers. While good solutions for the TSP can be approximated classically for problem sizes up to millions of cities, problems of similar structure like the CVRP are hard to solve to optimality even for hundreds of cities. In supply chain use cases, it is also important to achieve stable solutions. In addition to optimality, algorithms must achieve a good average across all evaluated solutions. Recent work has demonstrated that the Quantum Approximate Optimization Algorithm (QAOA)~\cite{farhi_quantum_2014} fails to solve the TSP as part of the CVRP when formulated with a one-hot encoding \cite{belly2023quantumassistedCVRP}. Even though the one-hot encoding is a very intuitive approach to formulating the TSP as an Ising model as described in \cite{Lucas_2014}, it leads to many Pauli-$Z$ eigenstates representing infeasible tours. Thus, finding good tours requires a very good performance of the quantum algorithm to approximation ratios above 99\% even for the smallest problem instances~\cite{belly2023quantumassistedCVRP}. The process can be simplified by reducing the search space. Instead of searching through and evaluating infeasible tours, we would like to restrict the search, ideally to a feasible subspace. To this end, the Alternating Operator Ansatz defines suitable mixers for instances of a given problem class \cite{hadfield_quantum_2019}. However, this requires a fault-tolerant quantum computer, since noisy mixer operators can produce infeasible states. Another approach is choosing a different encoding, where fewer bitstrings correspond to infeasible solutions. In recent years, several suggestions have been raised for encoding QUBO problems. Tan et al.~\cite{Tan_qubitEfficientEncoding_2021} propose a qubit-efficient encoding scheme for general QUBO problems, where an $n$-variable problem is reduced to $\mathcal{O}(\log n)$ variables. However, the suggested minimal encoding can only capture classical correlations and fails to find the correct configuration if the optimal solution is degenerate. A comparison of this qubit-efficient minimal encoding scheme to the one-hot encoding is investigated in Ref.~\cite{leonidas2023qubitEfficientRouting} for the vehicle routing problem with time windows. They observe a similar performance for both encodings, where the minimal encoding uses far less qubits. Unfortunately, the performance of QAOA with a one-hot-encoded TSP is already not good, so we do not expect an improvement for the minimal encoding. 
	In general, it is hard for an end user to decide a priori which encoding will fulfill their requirements the best.
	In this work, we investigate different encodings for the TSP and compare their scaling and performance for the QAOA and the Variational Quantum Eigensolver (VQE)~\cite{peruzzo_variational_2014}. We conclude with an analysis and comparison of all three encodings.

	\section{The TSP and its Encodings} 
	\label{sec: encodings}
	The TSP aims to find the shortest path in a weighted graph which visits each node exactly once and returns to its origin. In this work, we only consider fully connected graphs.
	
	When solving this task on a quantum computer, every possible route must be encoded with binary variables. The binary statevector can then be mapped onto qubits to minimize the distance depending on a quantum state. In the following, we investigate three different encodings.
	
	\subsection{TSP as QUBO}
	The most renowned encoding for the TSP results in a Quadratic Unconstrained Binary Optimization (QUBO) problem~\cite{Lucas_2014}. For $n$ nodes in the graph, each timestep $t \in \{1,..., n\}$ in the route uses a one-hot encoding to denote the index of the node visited. Therefore, the binary variable $x_{it}$ is the indicator for the statement ``Node $i$ is visited at time step $t$''. If the statement is true, it is $1$, otherwise $0$. To encode a route in a graph with $n$ nodes, this encoding requires a total of $n^2$ qubits. Given an encoding of a valid route, the length of the route to be minimized can then be evaluated via
	\begin{equation}
		C_d = \sum_{i=1}^n \sum_{j=1}^n w_{ij}\sum_{t=1}^n x_{it} x_{jt+1}
	\end{equation}
	with the distance between nodes $i$ and $j$ given as $w_{ij}$. Additionally, as a simplification, all indices here are considered $\mod n$. A disadvantage of this encoding is the possibility of obtaining solutions that do not correspond to a valid TSP route, so-called infeasible states. This is the case if either multiple cities are visited at the same timestep, i.e. the one-hot encoding contains multiple ones, or if the same city is visited multiple times in one route. To combat these infeasible solutions, the QUBO encoding introduces the penalty terms
	\begin{subequations}
		\begin{align}
			C_p^1 &= P \sum_{i = 1}^n \left(1 - \sum_{t=1}^n x_{it}\right)^2, \\
			C_p^2 &= P \sum_{t = 1}^n \left(1 - \sum_{i=1}^n x_{it}\right)^2
		\end{align}
	\end{subequations}
	with a penalty factor $P$. The subspace of infeasible states produced by this encoding is relatively large. Since the states contain $n^2$ qubits for one route and there are $n!$ possible routes, the subspace of feasible solutions in comparison to all states is $n!/2^{n^2}$. For increasing $n$, this expression decays superexponentially with $\mathcal{O}(2^{-n^2})$. 
	
	\subsection{TSP as HOBO}
	Another encoding strategy that reduces the number of necessary qubits and increases the size of the feasible subspace is the Higher Order Binary Optimization (HOBO) encoding \cite{glos_spaceefficientHOBO_2022}. Instead of denoting the (integer) label of a node at a timestep $t$ with one-hot encoding, this encoding represents the label by a binary number $b_t$. Each binary number $b_t$ contains $K \coloneqq \lceil \log(n) \rceil$ qubits, which results in $n \lceil \log n \rceil$ qubits necessary to encode a complete route. 
	For this encoding we first define $H_\mathrm{valid}$ which ensures that the binary index for a timestep does not exceed the number of nodes in the graph. Let $\tilde{b}_{K - 1} \dots \tilde{b}_0$ be a binary representation of $n - 1$ and define $k^0 \in K_0$ to be indices such that $\tilde{b}_{k^0} = 0$. Then the function $H_\mathrm{valid}$ takes the form
	\begin{equation}
		H_\mathrm{valid}(b_t) \coloneqq \sum_{k^0 \in K_0} b_{t,k^0} \prod_{k = k^0 + 1}^{K - 1} \left(1 - (b_{t,k} - \tilde{b}_k)^2\right).
	\end{equation}
	Further, the Hamiltonian term $H_\delta$ is constructed to compare two binary numbers, evaluating to one if they are equal and zero otherwise. This function determines the timesteps at which nodes are visited. All in all, $H_\delta$ is defined as
	\begin{equation}
		H_\delta(b, b') \coloneqq \prod_{k = 1}^K \left(1 - (b_k - b'_k)^2\right).
	\end{equation}
	The full Hamiltonian for the HOBO encoding is given by
	\begin{equation}
		\begin{split}
			C = P \sum_{t=1}^n H_\mathrm{valid}(b_t) + P \sum_{t=1}^n \sum_{t'=t + 1}^n H_\delta(b_t, b_{t'}) \\+ \sum_{i=1}^n \sum_{j=1}^n w_{ij} \sum_{t=1}^n H_\delta(b_t, i) H_\delta(b_{t + 1}, j).
		\end{split}
	\end{equation}
	Using this encoding, the portion of feasible states among all measurable states is given by $n! / 2^{n \lceil\log n \rceil}$. Unlike for QUBO, this ratio only decays exponentially of order $e^{-n}$. Thus, it is significantly larger but still approaches zero quickly with an increasing number of nodes.
	
	\subsection{Permutation Encoding for the TSP}
	The permutation encoding is designed to map each state to a valid TSP route. Glos et al. discuss this idea briefly in their work of 2022~\cite{glos_spaceefficientHOBO_2022}. In our work, we give an explicit implementation of this idea and provide numerical results. For this encoding, we identify each feasible solution with an integer number, the index, and interpret each measured state as a binary number corresponding to such an index of a TSP route. The routes are ordered such that they can be accessed efficiently. This way, we avoid enumerating and listing all possible paths. To achieve this, we interpret each TSP route as a permutation of the nodes. The permutations are ordered based on the generative Heap's Algorithm \cite{sedgewick1977permutation}. That means the order in which permutations are generated with Heap's Algorithm corresponds to the indices with which the permutations are accessed. The full procedure of mapping a binary state $b$ to a permutation is described in Algorithm~\ref{alg:perm}.
	\begin{table}[h]
		\begin{tabular}{c | c c c c} 
			\hline
			& \textbf{4 nodes} & \textbf{5 nodes} & \textbf{6 nodes} & \textbf{n nodes} \\ 
			\hline\hline
			\textbf{QUBO} & $3.66$ $10^{-4}$ & $3.57$ $10^{-6}$ & $1.04$ $10^{-8}$ & $\frac{n!}{2^{n^2}}$ \\ 
			\hline
			\textbf{HOBO} & $9.37$ $10^{-2}$ & $3.66$ $10^{-3}$ & $2.74$ $10^{-3}$ & $\frac{n!}{2^{n \lceil\log n \rceil}}$ \\
			\hline
			\textbf{Permutation} & $1$ & $1$ & $1$ & $1$ \\
			\hline
		\end{tabular}
		\caption{Ratio of the feasible subspace for TSP instances for increasing numbers of nodes using the described encodings. The HOBO encoding generates a larger ratio than the QUBO encoding. However, both ratios, QUBO and HOBO, approach zero for large problem instances,  whereas the permutation encoding only generates feasible states.}
		\label{table:feas_ratios}
	\end{table}
	With this mapping, every measured state corresponds to a valid TSP solution. Therefore, the size of the feasible subspace with regard to all states is increased to $1$. Additionally, this encoding uses fewer qubits than the QUBO and HOBO encoding with only $\lceil \log(n!) \rceil$ qubits for $n$ nodes. Yet, no efficient construction of a Hamiltonian representing the cost function is known for this encoding. The application of the permutation encoding is, hence, limited to methods where the Hamiltonian does not need to be known. In VQE, we can use a classical function to evaluate the measured bitstrings, instead of computing the expectation value of a Hamiltonian.
	\begin{algorithm}
		\caption{Mapping procedure for the permutation encoding. This algorithm receives a measured state $x$, which is interpreted as a binary number, and the total number of nodes $n$ as an input. The output is a permutation of the numbers $\{1, \dots, n\}$, which can be interpreted as a route in a TSP with $n$ nodes. The runtime for this algorithm is $\mathcal{O}(n)$.}\label{alg:perm}
		\KwData{Index $x$, number of nodes $n$}
		\KwResult{TSP Route $s$}
		$f \gets n!$, $y \gets x \mod f$\;
		$nodes \gets (1, \dots, n)$, $s \gets ()$, $i \gets 0$\;
		\While{$i < n$}{
			$f \gets \frac{f}{n - i}$\;
			$k \gets \lfloor \frac{x}{f} \rfloor$\;
			Append $nodes_k$ to $s$\;
			Delete $nodes_k$\;
			$x \gets x - kf$\;
			$i \gets i + 1$\;
		}
	\end{algorithm}
	
	\section{Variational Quantum Algorithms for Optimization}
	\label{sec:vqa}
	Variational Quantum Algorithms (VQA)~\cite{Cerezo_2021} are the leading proposal to approximate solutions to optimization problems with noisy intermediate-scale quantum (NISQ) devices~\cite{preskill_quantum_2018}. To avoid the accumulation of errors, they combine relatively small parameterized quantum circuits with a classical optimization algorithm that updates its parameters. Through quantum effects like entanglement and superposition, one hopes to find better approximations than classical algorithms, or find approximations of the same quality more efficiently.
	
	For a given combinatorial optimization problem, the candidate solutions need to be represented as binary vectors $\vec{x} \in \{0,1\}^n$ (encoding) that can then be mapped directly onto $n$-qubit quantum states $\ket{\Psi}$. The optimization goal is then represented as a cost function $C(\vec{x})$, or with slight abuse of notation $C(\ket{\Psi})$. A parameterized quantum circuit with unitary $U(\vec{\theta})$ depends on $s$ parameters $\vec{\theta} \in \mathbb{R}^s$, typically rotation angles. It is used to create quantum states via $\ket{\Psi} = U(\vec{\theta}) \ket{\Psi_0}$ from an initial state $\ket{\Psi_0}$, the standard choice being the uniform superposition of computational basis states. From the perspective of the classical optimizer, this is equivalent to a variable transformation of the cost function from binary vectors to the variational parameters. The task is then to minimize this function:
	\begin{equation}
		\min_{\vec{\theta}} C(\vec{\theta}) \coloneqq C(U(\vec{\theta}) \ket{\Psi_0})
	\end{equation}
	
	After the minimization has converged to optimized parameters $\vec{\theta}^*$, the quantum state $U(\vec{\theta}^*)\Psi_0$ can be measured in the computational basis to obtain a set of bitstrings that, interpreted as binary vectors, yield good solutions to the original optimization problem. The variational quantum eigensolver (VQE) implements this exact protocol while modelling $C(\vec{\theta})$ as the expectation value of a problem Hamiltonian $H_P$ (whose ground state is searched) as $C(\vec{\theta}) = \bra{\Psi_0} U^\dagger(\vec{\theta}) H_P U(\vec{\theta}) \ket{\Psi_0}$. However, in the context of combinatorial optimization, $H$ will be diagonal. Therefore, it does not need to be measured directly and can be calculated classically from the computational basis states in the measurement outcome. This eliminates the need to find an efficient implementation of the cost Hamiltonian. The last main ingredient for VQE is the ``ansatz'', the parameterized quantum circuit employed in the algorithm. For this study, we adopt a hardware-efficient ansatz designed to fit well onto real quantum hardware by using native gates and simple combinations of them (see \cref{sec:performance}).
	
	A special case of the VQE is the Quantum Approximate Optimization Algorithm (QAOA)~\cite{farhi_quantum_2014}. In the slightly more general version as Quantum Alternating Operator Ansatz~\cite{hadfield_quantum_2019}, it is defined by the problem Hamiltonian $H_P$ encoding the optimization problem and a mixer Hamiltonian $H_M$. For the original QAOA, $H_M$ is the x-mixer $\sum_{j=1}^n \hat{\sigma}^x_j$. The initial quantum state is then evolved alternatingly with $H_P$ and $H_M$, with the evolution times typically denoted by $\gamma_j$ and $\beta_j$ respectively. Intuitively, $H_P$ applies a quality-dependent phase shift, whereas $H_M$ mixes the computational basis states. The full ansatz reads
	\begin{align}
		U(\vec{\beta}, \vec{\gamma}) = \prod_{j=1}^p e^{-i\beta_j H_M}  e^{-i\gamma_j H_P}.
	\end{align}
	
	The interest in QAOA has been sparked by analytical performance guarantees~\cite{farhi_quantum_2014} for special cases, although it remains an open question whether an efficient training of the ansatz is possible and how good initial points can be found~\cite{rajakumar_trainability_2024, bravyi_obstacles_2020, egger_warm-starting_2021}. Furthermore, QAOA can be seen as a discretized version of adiabatic quantum computing with $\beta_j$ decreasing through the layers while the $\gamma_j$ values increase.
	
	\section{Performance of different Encodings}
	\label{sec:performance}
	
	The performance of VQE and QAOA on the TSP is evaluated numerically for all three encoding strategies using simulated circuits during the optimization. Specifically, we use Qiskit Aer's Statevector Simulator to perform the circuit evaluations. As performance measures, we use the feasibility ratio and the TSP length ratio according to Ref.~\cite{belly2023quantumassistedCVRP}. The feasibility ratio $r_f$ corresponds to the percentage of feasible measured shots with the formula
	\begin{equation}
		r_f = \frac{\text{\#feasible shots}}{\text{\#total shots}}.
	\end{equation}
	The ratio lies between $0$ (corresponding to no feasible shots) and $1$ (only feasible shots). For the TSP length ratio $r_\ell$, we introduce a modified version from Ref.~\cite{belly2023quantumassistedCVRP} to achieve consistent lower and upper bounds of $0$ and $1$ across different instances. Let
	\begin{subequations}
		\begin{align}
			\ell_{\text{min}} \coloneqq \min_{s \in S_f} \ell(s)\\
			\ell_{\text{max}} \coloneqq \max_{s \in S_f} \ell(s)
		\end{align}
	\end{subequations}
	be the minimal and maximal length of a TSP instance. Here, $S_f$ denotes the set of all feasible states, and $\ell(s)$ is the TSP length of the route encoded by state $s$. Then 
	\begin{equation}
		\tilde{r}_\ell = \frac{\sum_{s \in S_f} p(s) \frac{\ell_{\text{min}}}{\ell(s)}}{r_f}
	\end{equation}
	defines the length ratio with $p(s)$ being the probability of sampling a state $s$. This ratio has an upper bound of $1$ if the optimal route is always achieved and a lower bound of
	\begin{equation}
		\min_{\tilde{r}_\ell} = \frac{\ell_{\text{min}}}{\ell_{\text{max}}}
	\end{equation}
	if only the worst solution is found. Since this lower bound depends on the specific TSP instances, we rescale it linearly to be in the range between $0$ and $1$, resulting in
	\begin{equation}
		r_\ell = \frac{\tilde{r}_\ell - \min_{\tilde{r}_\ell}}{1-\min_{\tilde{r}_\ell}}.
	\end{equation}
	This ensures sensible comparability of the quality of the found solutions between different instances and TSP sizes. If the feasibility ratio $r_f$ is $0$, we also define the length ratio to take the minimum value of $0$. 
	
	For VQE, we use a two-local circuit with alternating rotation and entanglement layers. The first layer consists of single-qubit $Y$-rotation gates with the rotation angles as adjustable parameters. It is followed by a circular entanglement layer, where a two-qubit CX-gate is applied to each pair of neighboring qubits. This ansatz is repeated once, and an additional rotational layer is added at the end to create five layers with $3n$ parameters, with $n$ being the number of qubits. In total, this ansatz has a depth that scales linearly with the number of qubits.
	
	Additionally, we evaluate the performance of the encodings using QAOA with depth $p=2$. The mixer is chosen to be the sum of Pauli-$X$ Gates. Since we cannot access an efficiently constructed Hamiltonian for the permutation encoding, QAOA can only be used in conjunction with the QUBO and HOBO encoding.
	
	We test the performance on TSP instances of 4, 5, and 6 nodes, where the first city is always fixed for each solution path to reduce the problem's dimension. The coordinates of the nodes are drawn from a uniform random distribution in a $100\times 100$ box. All instances are formulated with a penalty of $P=100$ for the QUBO encoding and $P=200$ for the HOBO encoding. We chose a base penalty of the maximum possible distance between two nodes to encourage the optimizer to select a feasible solution while not disregarding the TSP length of feasible solutions. For the HOBO encoding, the penalty needs to be increased since the minimum energy eigenstate can be infeasible otherwise. We test both gradient-based optimizers like SLSQP, SPSA, and CG, as well as gradient-free optimizers such as NFT, Powell, COBYLA, Nelder-Mead, and UMDA. Apart from the last one, all optimizers are local. The performance measures $r_\ell$ and $r_f$ for one concrete instance of each TSP size can be seen in Figure~\ref{fig:ratios}. For each instance, the mean value and standard deviation of 40 measurements with different initial parameter values is reported. The initial rotation angles are drawn randomly from a uniform distribution within $[0, 2\pi]$. Additional experiments showed that different instances of the same TSP size result in a relatively stable performance, i.e., the occurring fluctuations are comparable to those for varying starting points. Therefore, we focus on one specific TSP instance for each problem size.
	\begin{figure}[h]
		\centering
		\begin{subfigure}{0.5\textwidth}
			\captionsetup{singlelinecheck = false, format= hang, justification=raggedright}
			\caption{}
			\input{tikz_figures/rescaled_lr}
			\label{fig:lr}
		\end{subfigure}
		\begin{subfigure}{0.5\textwidth}
			\captionsetup{singlelinecheck = false, format= hang, justification=raggedright}
			\caption{}
			\input{tikz_figures/fr}
			\label{fig:fr}
		\end{subfigure}
		\caption{(a) Rescaled Length Ratio $r_\ell$ and (b) Feasibility Ratio $r_f$ for TSP instances with up to $6$ nodes. The tested encodings are the QUBO, HOBO, and permutation encoding with VQE, and the former two encodings with QAOA. The optimization is done with 40 initial parameters from which the average and standard deviation of the results are evaluated. Each method is trained with the NFT optimizer and uses the same TSP instance.}
		\label{fig:ratios}
	\end{figure}
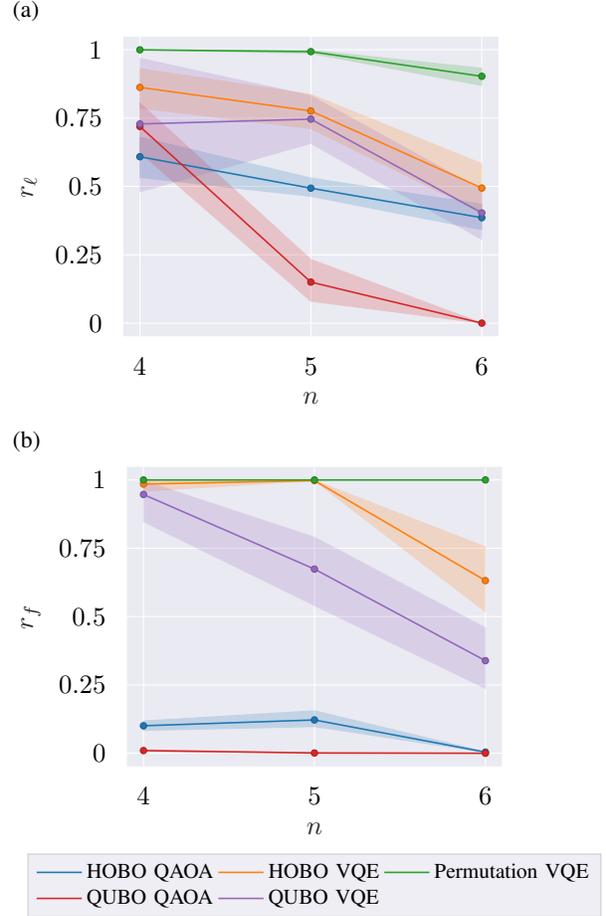
	\begin{figure}
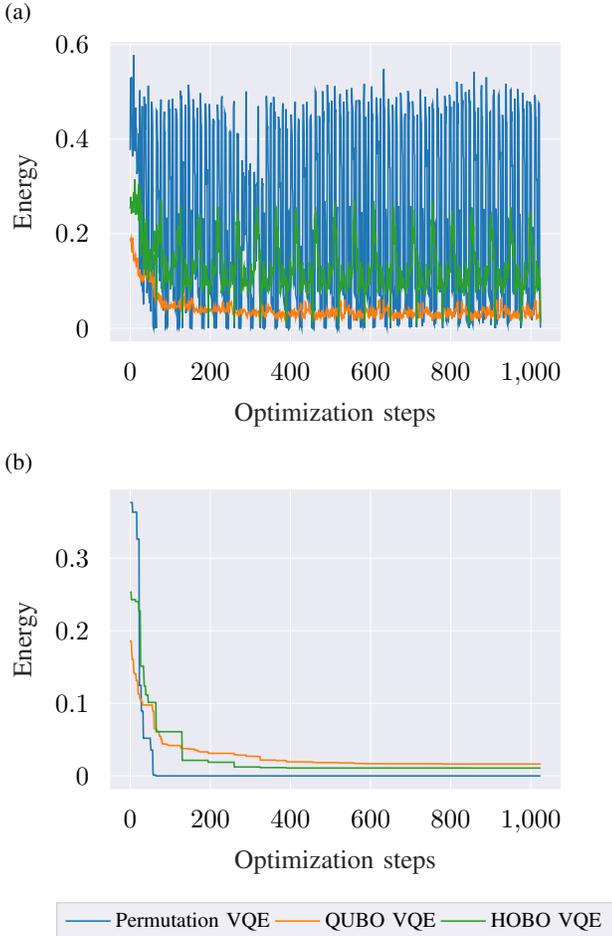

		\centering
		\begin{subfigure}{0.5\textwidth}
			\captionsetup{singlelinecheck = false, format= hang, justification=raggedright}
			\caption{}
			\input{tikz_figures/optimization}
			\label{fig:opt_total}
		\end{subfigure}
		\begin{subfigure}{0.5\textwidth}
			\captionsetup{singlelinecheck = false, format= hang, justification=raggedright}
			\caption{}
			\input{tikz_figures/optimization_min}
			\label{fig:opt_min}
		\end{subfigure}
		\caption{Energies during optimization with NFT of the three encodings using VQE for a TSP instance with $5$ nodes. (a) shows all evaluated energies during the optimization and (b) the moving minimum of the same energies. To allow for better comparison, the energies are linearly re-scaled such that $0$ represents the minimum achievable energy while $1$ represents the worst case energy.}
		\label{fig:enter-label}
	\end{figure}
	With increasing problem size, the feasibility ratio of the QUBO and the HOBO encoding falls off significantly (Figure \ref{fig:fr}). While the HOBO encoding still achieves a near-perfect feasibility ratio for $5$ nodes, that ratio drops for both the QUBO encoding and the HOBO encoding for 6 nodes. Meanwhile, the permutation encoding translates every state into a feasible TSP solution and achieves a constant feasibility ratio of $1$, as expected.
	\begin{figure*}
		\centering
		\tikzset{every picture/.style={scale=0.66}}
		\begin{subfigure}{0.26\textwidth}
			\captionsetup{singlelinecheck = false, format= hang, justification=raggedright}
			\caption{VQE with QUBO}
			\input{tikz_figures/qubo_landscape}
		\end{subfigure}
		\begin{subfigure}{0.26\textwidth}
			\captionsetup{singlelinecheck = false, format= hang, justification=raggedright}
			\caption{VQE with HOBO}
			\input{tikz_figures/hobo_landscape}
		\end{subfigure}
		\begin{subfigure}{0.26\textwidth}
			\captionsetup{singlelinecheck = false, format= hang, justification=raggedright}
			\caption{VQE with permutation encoding}
			\input{tikz_figures/permutation_landscape}
		\end{subfigure}
		\begin{subfigure}{0.26\textwidth}
			\captionsetup{singlelinecheck = false, format= hang, justification=raggedright}
			\caption{QAOA with QUBO}
			\input{tikz_figures/qubo_qaoa_landscape}
		\end{subfigure}
		\begin{subfigure}{0.26\textwidth}
			\captionsetup{singlelinecheck = false, format= hang, justification=raggedright}
			\caption{QAOA with HOBO}
			\input{tikz_figures/hobo_qaoa_landscape}
		\end{subfigure}
		\caption{Energy landscape for all encodings when adjusting randomly chosen parameters $\theta_0$ and $\theta_1$ in the parameter space of the corresponding ansatz. 
			All methods are evaluated for the same TSP instance with $4$ nodes and the same adapted parameters $\theta_0$ and $\theta_1$.}
		\label{fig:landscapes}
	\end{figure*}
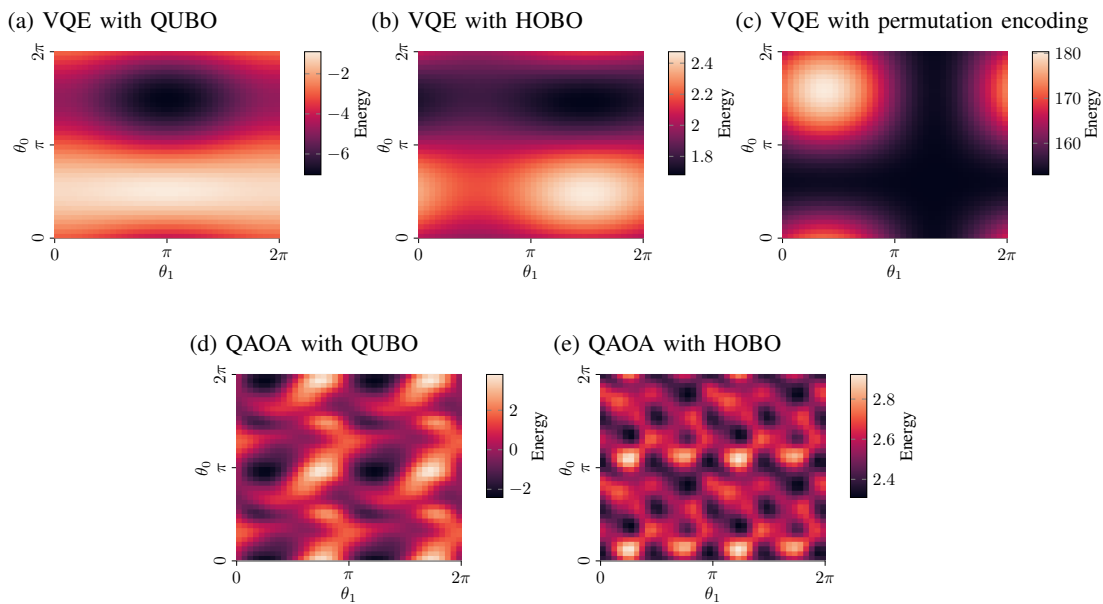
	
	As seen in Figure~\ref{fig:lr}, using the HOBO encoding leads to lower length ratios than those found with the QUBO encoding. Yet, the permutation encoding outperforms both of these methods, finding solutions significantly closer to the optimal paths. Also, the permutation encoding performs more consistently than the other encoding, achieving a lower standard deviation than both the QUBO and the HOBO encoding.
	
	Since we not only want to achieve optimality but also good averages, we also investigate the energies for each method. To generate energy landscapes, we chose two random parameters of the ansatz and evolved these values between $0$ and $2 \pi$. For the remaining parameters, the values resulting from the optimization are retained. Then, we execute the circuit and evaluate the energy with respect to the corresponding Hamiltonian for each method. Since the permutation encoding does not have a Hamiltonian, we measure the TSP length of each state weighted by the probability of sampling that state after running the circuit. Similarly to the other results, we do not use the permutation encoding in combination with QAOA due to the absent Hamiltonian. When investigating the resulting landscapes shown in Figure~\ref{fig:landscapes}, it becomes clear that the energy landscapes depend more on the ansatz circuit than the chosen encoding. While the VQE circuits produce fairly simple landscapes that seem very easy to optimize, the QAOA circuits are more complex, with multiple local minima where an optimizer might get trapped in. In this case, QAOA does not have a guarantee for a periodicity of $2 \pi$ since the eigenvalues of the Hamiltonian are not always integers. Yet, it is still possible to observe a repeating pattern with a periodicity of $\pi$ in the landscapes.
	
	When investigating the energies achieved during optimization (see Figure~\ref{fig:enter-label}), the optimizer converges faster using the permutation encoding than with both other methods. However, the oscillations during optimization with this encoding are significantly larger. Additionally, the optimizer does not reach a clear convergence and fails to optimize the energy further after about $50$ steps. While this might be related to the optimizer trying to cover a large space of parameters, it might also point towards the optimization for this encoding being harder than the other optimizations. Considering the small number of nodes of the instances tested here, the optimizer might have been able to find a solution by randomly sampling TSP paths instead of actually capturing the properties. To investigate this possibility, it would be interesting to explore the scaling of the permutation encoding for larger instances in further work. While the optimization curve of the HOBO encoding also shows larger oscillations than with the QUBO encoding, this effect is much smaller than with the permutation encoding. The converged energies of the methods seem to align with their performance tested previously, with the HOBO encoding performing better than the QUBO encoding and the permutation encoding outperforming both other methods.
	
	\section{Conclusions}
	\label{sec:conclusions}
	In this work, we investigated three different encodings for the TSP and compared their performance with regard to the feasibility and optimality of their solutions. The different encodings were implemented, analyzed in theory and tested numerically on simulators. The permutation encoding strongly benefits from its inability to create infeasible solutions, outperforming both the QUBO and HUBO encoding. It produces more feasible solutions and solutions that are closer to the optimum compared to the other two encodings. However, there is no known efficient calculation of a Hamiltonian for the permutation encoding. This drawback limits the method to being used alongside algorithms like VQE, where no Hamiltonian is necessary. Additionally, it seems that the classical optimizer struggles to identify geometric properties of the permutation encoding, resulting in a noisy optimization. Even though this optimizer still often finds the optimal solution and outperforms the other encodings by means of feasibility and length of the solution, it might be interesting to investigate the permutation encoding for larger instances in the future. In a TSP with 6 nodes, there are only 120 paths, of which 2 are optimal. Therefore, for this size of a problem, it would still be feasible to randomly sample solutions and come across the optimal one. To look further into this possibility, it would be necessary to test the performance of the permutation encoding for larger instances. Generally it is still an open question whether VQE and QAOA are suitable algorithms for solving combinatorial optimization problems, particularly for larger instances.
	
	The QUBO encoding is probably the most intuitive approach. In contrast to HOBO, the QUBO encoding produces only quadratic cost functions. The resulting quantum circuits require shorter depth when compiled to two-qubit gates compared to cost functions of higher order. However, the HOBO encoding requires fewer qubits than the QUBO encoding. Since the HOBO encoding outperforms the QUBO in terms of the feasibility ratio and quality of the solutions, we conclude that HOBO is more advisable than QUBO encoding whenever the available quantum hardware can run a circuit of the required depth.
	
	When comparing encoding strategies, there seems to be a clear trade-off between the complexity of the cost function and the algorithm's performance, with more complex cost functions leading to better results. Ultimately, this might severely limit the capacities of more complex encodings on real hardware (e.g., implementation of a HOBO Hamiltonian). 
	
	Generally speaking, based on the results we obtained for instances up to 6 nodes, we recommend the permutation encoding of the TSP due to its clear advantages in performance. Even in the presence of errors, the permutation encoding will always produce feasible solutions. However, it remains an open question whether the good performance of the permutation encoding is due to the small problem size. In general, there is no performance guarantee for solving the TSP for any encoding with VQE. Therefore, the quality of solutions for larger instances is unknown. In a setting where a Hamiltonian is necessary for the optimization, we recommend the HOBO encoding over the QUBO encoding since it produces better solutions with fewer qubits.

	\section*{Acknowledgement}
	This project is supported by the Federal Ministry for Economic Affairs and Climate Action on the basis of a decision by the German Bundestag through the project \emph{Quantum-enabling Services and Tools for Industrial Applications (QuaST)}. QuaST aims to facilitate the access to quantum-based solutions for optimization problems and to bridge the gap between business and technology. Furthermore, we would like to thank Philipp Seitz for his valuable comments.

	\bibliographystyle{unsrt}
	\balance
	\bibliography{main.bib}
\end{document}

%% file: tikz_figures/rescaled_lr.tex
% This file was created with tikzplotlib v0.10.1.
\begin{tikzpicture}

\definecolor{crimson2143940}{RGB}{214,39,40}
\definecolor{darkorange25512714}{RGB}{255,127,14}
\definecolor{darkslategray38}{RGB}{38,38,38}
\definecolor{forestgreen4416044}{RGB}{44,160,44}
\definecolor{lavender234234242}{RGB}{234,234,242}
\definecolor{lightgray204}{RGB}{204,204,204}
\definecolor{mediumpurple148103189}{RGB}{148,103,189}
\definecolor{steelblue31119180}{RGB}{31,119,180}

\begin{axis}[
axis background/.style={fill=lavender234234242},
axis line style={white},
legend cell align={left},scale only axis, width=5cm,height=4cm,
legend columns=3,
legend style={
nodes={scale=0.8, transform shape},
  fill opacity=0.8,
  draw opacity=1,
  text opacity=1,
  at={(0.5,-0.35)},
  anchor=south,
  draw=lightgray204,
  fill=lavender234234242
},
tick align=outside,
x grid style={white},
xlabel=\textcolor{darkslategray38}{\(\displaystyle n\)},
xmajorgrids,
xmin=3.9, xmax=6.1,
xtick style={draw=none},
ytick style={draw=none},
xtick={4, 5, 6},
ytick={0.0, 0.25, 0.5, 0.75, 1.0},
xticklabels={$4$,$5$,$6$},
xtick style={color=darkslategray38},
y grid style={white},
ylabel=\textcolor{darkslategray38}{\(\displaystyle r_\ell\)},
ymajorgrids,
ymajorticks=true,
ymin=-0.0499798882409392, ymax=1.04957765305972,
ytick style={color=darkslategray38}
]
\path [draw=steelblue31119180, fill=steelblue31119180, opacity=0.2]
(axis cs:4,0.679953689851797)
--(axis cs:4,0.532946595036274)
--(axis cs:5,0.463850332755423)
--(axis cs:6,0.34171464023551)
--(axis cs:6,0.434903307822576)
--(axis cs:6,0.434903307822576)
--(axis cs:5,0.530572426697348)
--(axis cs:4,0.679953689851797)
--cycle;

\path [draw=darkorange25512714, fill=darkorange25512714, opacity=0.2]
(axis cs:4,0.931009946033495)
--(axis cs:4,0.786550751477929)
--(axis cs:5,0.711753660372201)
--(axis cs:6,0.40349449120717)
--(axis cs:6,0.583226361773535)
--(axis cs:6,0.583226361773535)
--(axis cs:5,0.836376395211295)
--(axis cs:4,0.931009946033495)
--cycle;

\path [draw=forestgreen4416044, fill=forestgreen4416044, opacity=0.2]
(axis cs:4,0.999597764818785)
--(axis cs:4,0.999376945677452)
--(axis cs:5,0.986922015265633)
--(axis cs:6,0.869149162758424)
--(axis cs:6,0.932141744081173)
--(axis cs:6,0.932141744081173)
--(axis cs:5,0.997977249003886)
--(axis cs:4,0.999597764818785)
--cycle;

\path [draw=crimson2143940, fill=crimson2143940, opacity=0.2]
(axis cs:4,0.804688181421475)
--(axis cs:4,0.624701147447115)
--(axis cs:5,0.0803346049196662)
--(axis cs:6,0)
--(axis cs:6,0)
--(axis cs:6,0)
--(axis cs:5,0.232817172682851)
--(axis cs:4,0.804688181421475)
--cycle;

\path [draw=mediumpurple148103189, fill=mediumpurple148103189, opacity=0.2]
(axis cs:4,0.968308247270557)
--(axis cs:4,0.480495680079404)
--(axis cs:5,0.657393947299104)
--(axis cs:6,0.306416857338633)
--(axis cs:6,0.486861562934396)
--(axis cs:6,0.486861562934396)
--(axis cs:5,0.831681408393177)
--(axis cs:4,0.968308247270557)
--cycle;

\addplot [
  forget plot,
  mark=*,
  mark options={scale=0.6},
  only marks,
  scatter,
  scatter/@pre marker code/.code={%
            \edef\temp{\noexpand\definecolor{mapped color}{RGB}{\pgfplotspointmeta}}%
            \temp
            \scope[draw=mapped color!80!black,fill=mapped color]%
        },%
        scatter/@post marker code/.code={%
            \endscope
        },%
        point meta={TeX code symbolic={%
            \edef\pgfplotspointmeta{\thisrow{RED},\thisrow{GREEN},\thisrow{BLUE}}%
        }},
]
table{
x  y  RED GREEN BLUE
4 0.608549340465671 31 119 180
5 0.493553411852933 31 119 180
6 0.38566486689096 31 119 180
4 0.862286424380061 255 127 14
5 0.776111378436363 255 127 14
6 0.494051407042528 255 127 14
4 0.999496023058142 44 160 44
5 0.992956145087522 44 160 44
6 0.902737237974506 44 160 44
4 0.719264503048735 214 39 40
5 0.150066308595224 214 39 40
6 0 214 39 40
4 0.728340708331467 148 103 189
5 0.746045060275806 148 103 189
6 0.403090560954304 148 103 189
};

\addplot [semithick, steelblue31119180]
table {%
4 0.608549340465671
5 0.493553411852933
6 0.38566486689096
};

\addplot [semithick, darkorange25512714]
table {%
4 0.862286424380061
5 0.776111378436363
6 0.494051407042528
};
\addplot [semithick, forestgreen4416044]
table {%
4 0.999496023058142
5 0.992956145087522
6 0.902737237974506
};
\addplot [semithick, crimson2143940]
table {%
4 0.719264503048735
5 0.150066308595224
6 0
};
\addplot [semithick, mediumpurple148103189]
table {%
4 0.728340708331467
5 0.746045060275806
6 0.403090560954304
};
%\legend{HOBO QAOA, HOBO VQE, Permutation VQE, QUBO QAOA, QUBO VQE}
\end{axis}

\end{tikzpicture}

%% file: tikz_figures/fr.tex
% This file was created with tikzplotlib v0.10.1.
\begin{tikzpicture}

\definecolor{crimson2143940}{RGB}{214,39,40}
\definecolor{darkorange25512714}{RGB}{255,127,14}
\definecolor{darkslategray38}{RGB}{38,38,38}
\definecolor{forestgreen4416044}{RGB}{44,160,44}
\definecolor{lavender234234242}{RGB}{234,234,242}
\definecolor{lightgray204}{RGB}{204,204,204}
\definecolor{mediumpurple148103189}{RGB}{148,103,189}
\definecolor{steelblue31119180}{RGB}{31,119,180}

\begin{axis}[
axis background/.style={fill=lavender234234242},
axis line style={white},
legend cell align={left},scale only axis, width=5cm,height=4cm,
legend columns=3,
legend style={
nodes={scale=0.8, transform shape},
  fill opacity=0.8,
  draw opacity=1,
  text opacity=1,
  at={(0.5,-0.5)},
  anchor=south,
  draw=lightgray204,
  fill=lavender234234242
},
tick align=outside,
x grid style={white},
xlabel=\textcolor{darkslategray38}{\(\displaystyle n\)},
xmajorgrids,
xmin=3.9, xmax=6.1,
xtick style={draw=none},
ytick style={draw=none},
xtick={4, 5, 6},
ytick={0.0, 0.25, 0.5, 0.75, 1.0},
xticklabels={$4$,$5$,$6$},
xtick style={color=darkslategray38},
y grid style={white},
ylabel=\textcolor{darkslategray38}{\(\displaystyle r_f\)},
ymajorgrids,
ymajorticks=true,
ymin=-0.05, ymax=1.05,
ytick style={color=darkslategray38}
]
\path [draw=steelblue31119180, fill=steelblue31119180, opacity=0.2]
(axis cs:4,0.11807373046875)
--(axis cs:4,0.0835675048828125)
--(axis cs:5,0.0976519775390625)
--(axis cs:6,0.003466796875)
--(axis cs:6,0.0048583984375)
--(axis cs:6,0.0048583984375)
--(axis cs:5,0.155350341796875)
--(axis cs:4,0.11807373046875)
--cycle;

\path [draw=darkorange25512714, fill=darkorange25512714, opacity=0.2]
(axis cs:4,0.998145141601563)
--(axis cs:4,0.959033203125)
--(axis cs:5,0.9970947265625)
--(axis cs:6,0.518135375976563)
--(axis cs:6,0.75490966796875)
--(axis cs:6,0.75490966796875)
--(axis cs:5,0.998193359375)
--(axis cs:4,0.998145141601563)
--cycle;

\path [draw=forestgreen4416044, fill=forestgreen4416044, opacity=0.2]
(axis cs:4,1)
--(axis cs:4,1)
--(axis cs:5,1)
--(axis cs:6,1)
--(axis cs:6,1)
--(axis cs:6,1)
--(axis cs:5,1)
--(axis cs:4,1)
--cycle;

\path [draw=crimson2143940, fill=crimson2143940, opacity=0.2]
(axis cs:4,0.012672119140625)
--(axis cs:4,0.007200927734375)
--(axis cs:5,0.0003173828125)
--(axis cs:6,0)
--(axis cs:6,0)
--(axis cs:6,0)
--(axis cs:5,0.0021734619140625)
--(axis cs:4,0.012672119140625)
--cycle;

\path [draw=mediumpurple148103189, fill=mediumpurple148103189, opacity=0.2]
(axis cs:4,0.99676878006275)
--(axis cs:4,0.84749818232425)
--(axis cs:5,0.541024209612687)
--(axis cs:6,0.237496337890625)
--(axis cs:6,0.458936767578125)
--(axis cs:6,0.458936767578125)
--(axis cs:5,0.790249571457937)
--(axis cs:4,0.99676878006275)
--cycle;

\addplot [
  forget plot,
  mark=*,
  mark options={scale=0.6},
  only marks,
  scatter,
  scatter/@pre marker code/.code={%
            \edef\temp{\noexpand\definecolor{mapped color}{RGB}{\pgfplotspointmeta}}%
            \temp
            \scope[draw=mapped color!80!black,fill=mapped color]%
        },%
        scatter/@post marker code/.code={%
            \endscope
        },%
        point meta={TeX code symbolic={%
            \edef\pgfplotspointmeta{\thisrow{RED},\thisrow{GREEN},\thisrow{BLUE}}%
        }},
]
table{%
x  y  RED GREEN BLUE
4 0.1005126953125 31 119 180
5 0.121923828125 31 119 180
6 0.0041748046875 31 119 180
4 0.9848876953125 255 127 14
5 0.9976806640625 255 127 14
6 0.631689453125 255 127 14
4 1 44 160 44
5 1 44 160 44
6 1 44 160 44
4 0.0098388671875 214 39 40
5 0.0010009765625 214 39 40
6 0 214 39 40
4 0.94685913559 148 103 189
5 0.67393749503 148 103 189
6 0.33876953125 148 103 189
};
\addplot [semithick, steelblue31119180]
table {%
4 0.1005126953125
5 0.121923828125
6 0.0041748046875
};
\addplot [semithick, darkorange25512714]
table {%
4 0.9848876953125
5 0.9976806640625
6 0.631689453125
};
\addplot [semithick, forestgreen4416044]
table {%
4 1
5 1
6 1
};
\addplot [semithick, crimson2143940]
table {%
4 0.0098388671875
5 0.0010009765625
6 0
};
\addplot [semithick, mediumpurple148103189]
table {%
4 0.94685913559
5 0.67393749503
6 0.33876953125
};
\legend{HOBO QAOA, HOBO VQE, Permutation VQE, QUBO QAOA, QUBO VQE}
\end{axis}

\end{tikzpicture}

%% file: tikz_figures/qubo_landscape.tex
% This file was created with tikzplotlib v0.10.1.
\begin{tikzpicture}

\definecolor{darkgray176}{RGB}{176,176,176}

\begin{axis}[
colorbar,
colorbar style={ylabel={Energy}, ylabel style={yshift=-0.1cm}},
colormap={mymap}{[1pt]
 rgb(0pt)=(0.01060815,0.01808215,0.10018654);
  rgb(1pt)=(0.01428972,0.02048237,0.10374486);
  rgb(2pt)=(0.01831941,0.0229766,0.10738511);
  rgb(3pt)=(0.02275049,0.02554464,0.11108639);
  rgb(4pt)=(0.02759119,0.02818316,0.11483751);
  rgb(5pt)=(0.03285175,0.03088792,0.11863035);
  rgb(6pt)=(0.03853466,0.03365771,0.12245873);
  rgb(7pt)=(0.04447016,0.03648425,0.12631831);
  rgb(8pt)=(0.05032105,0.03936808,0.13020508);
  rgb(9pt)=(0.05611171,0.04224835,0.13411624);
  rgb(10pt)=(0.0618531,0.04504866,0.13804929);
  rgb(11pt)=(0.06755457,0.04778179,0.14200206);
  rgb(12pt)=(0.0732236,0.05045047,0.14597263);
  rgb(13pt)=(0.0788708,0.05305461,0.14995981);
  rgb(14pt)=(0.08450105,0.05559631,0.15396203);
  rgb(15pt)=(0.09011319,0.05808059,0.15797687);
  rgb(16pt)=(0.09572396,0.06050127,0.16200507);
  rgb(17pt)=(0.10132312,0.06286782,0.16604287);
  rgb(18pt)=(0.10692823,0.06517224,0.17009175);
  rgb(19pt)=(0.1125315,0.06742194,0.17414848);
  rgb(20pt)=(0.11813947,0.06961499,0.17821272);
  rgb(21pt)=(0.12375803,0.07174938,0.18228425);
  rgb(22pt)=(0.12938228,0.07383015,0.18636053);
  rgb(23pt)=(0.13501631,0.07585609,0.19044109);
  rgb(24pt)=(0.14066867,0.0778224,0.19452676);
  rgb(25pt)=(0.14633406,0.07973393,0.1986151);
  rgb(26pt)=(0.15201338,0.08159108,0.20270523);
  rgb(27pt)=(0.15770877,0.08339312,0.20679668);
  rgb(28pt)=(0.16342174,0.0851396,0.21088893);
  rgb(29pt)=(0.16915387,0.08682996,0.21498104);
  rgb(30pt)=(0.17489524,0.08848235,0.2190294);
  rgb(31pt)=(0.18065495,0.09009031,0.22303512);
  rgb(32pt)=(0.18643324,0.09165431,0.22699705);
  rgb(33pt)=(0.19223028,0.09317479,0.23091409);
  rgb(34pt)=(0.19804623,0.09465217,0.23478512);
  rgb(35pt)=(0.20388117,0.09608689,0.23860907);
  rgb(36pt)=(0.20973515,0.09747934,0.24238489);
  rgb(37pt)=(0.21560818,0.09882993,0.24611154);
  rgb(38pt)=(0.22150014,0.10013944,0.2497868);
  rgb(39pt)=(0.22741085,0.10140876,0.25340813);
  rgb(40pt)=(0.23334047,0.10263737,0.25697736);
  rgb(41pt)=(0.23928891,0.10382562,0.2604936);
  rgb(42pt)=(0.24525608,0.10497384,0.26395596);
  rgb(43pt)=(0.25124182,0.10608236,0.26736359);
  rgb(44pt)=(0.25724602,0.10715148,0.27071569);
  rgb(45pt)=(0.26326851,0.1081815,0.27401148);
  rgb(46pt)=(0.26930915,0.1091727,0.2772502);
  rgb(47pt)=(0.27536766,0.11012568,0.28043021);
  rgb(48pt)=(0.28144375,0.11104133,0.2835489);
  rgb(49pt)=(0.2875374,0.11191896,0.28660853);
  rgb(50pt)=(0.29364846,0.11275876,0.2896085);
  rgb(51pt)=(0.29977678,0.11356089,0.29254823);
  rgb(52pt)=(0.30592213,0.11432553,0.29542718);
  rgb(53pt)=(0.31208435,0.11505284,0.29824485);
  rgb(54pt)=(0.31826327,0.1157429,0.30100076);
  rgb(55pt)=(0.32445869,0.11639585,0.30369448);
  rgb(56pt)=(0.33067031,0.11701189,0.30632563);
  rgb(57pt)=(0.33689808,0.11759095,0.3088938);
  rgb(58pt)=(0.34314168,0.11813362,0.31139721);
  rgb(59pt)=(0.34940101,0.11863987,0.3138355);
  rgb(60pt)=(0.355676,0.11910909,0.31620996);
  rgb(61pt)=(0.36196644,0.1195413,0.31852037);
  rgb(62pt)=(0.36827206,0.11993653,0.32076656);
  rgb(63pt)=(0.37459292,0.12029443,0.32294825);
  rgb(64pt)=(0.38092887,0.12061482,0.32506528);
  rgb(65pt)=(0.38727975,0.12089756,0.3271175);
  rgb(66pt)=(0.39364518,0.12114272,0.32910494);
  rgb(67pt)=(0.40002537,0.12134964,0.33102734);
  rgb(68pt)=(0.40642019,0.12151801,0.33288464);
  rgb(69pt)=(0.41282936,0.12164769,0.33467689);
  rgb(70pt)=(0.41925278,0.12173833,0.33640407);
  rgb(71pt)=(0.42569057,0.12178916,0.33806605);
  rgb(72pt)=(0.43214263,0.12179973,0.33966284);
  rgb(73pt)=(0.43860848,0.12177004,0.34119475);
  rgb(74pt)=(0.44508855,0.12169883,0.34266151);
  rgb(75pt)=(0.45158266,0.12158557,0.34406324);
  rgb(76pt)=(0.45809049,0.12142996,0.34540024);
  rgb(77pt)=(0.46461238,0.12123063,0.34667231);
  rgb(78pt)=(0.47114798,0.12098721,0.34787978);
  rgb(79pt)=(0.47769736,0.12069864,0.34902273);
  rgb(80pt)=(0.48426077,0.12036349,0.35010104);
  rgb(81pt)=(0.49083761,0.11998161,0.35111537);
  rgb(82pt)=(0.49742847,0.11955087,0.35206533);
  rgb(83pt)=(0.50403286,0.11907081,0.35295152);
  rgb(84pt)=(0.51065109,0.11853959,0.35377385);
  rgb(85pt)=(0.51728314,0.1179558,0.35453252);
  rgb(86pt)=(0.52392883,0.11731817,0.35522789);
  rgb(87pt)=(0.53058853,0.11662445,0.35585982);
  rgb(88pt)=(0.53726173,0.11587369,0.35642903);
  rgb(89pt)=(0.54394898,0.11506307,0.35693521);
  rgb(90pt)=(0.5506426,0.11420757,0.35737863);
  rgb(91pt)=(0.55734473,0.11330456,0.35775059);
  rgb(92pt)=(0.56405586,0.11235265,0.35804813);
  rgb(93pt)=(0.57077365,0.11135597,0.35827146);
  rgb(94pt)=(0.5774991,0.11031233,0.35841679);
  rgb(95pt)=(0.58422945,0.10922707,0.35848469);
  rgb(96pt)=(0.59096382,0.10810205,0.35847347);
  rgb(97pt)=(0.59770215,0.10693774,0.35838029);
  rgb(98pt)=(0.60444226,0.10573912,0.35820487);
  rgb(99pt)=(0.61118304,0.10450943,0.35794557);
  rgb(100pt)=(0.61792306,0.10325288,0.35760108);
  rgb(101pt)=(0.62466162,0.10197244,0.35716891);
  rgb(102pt)=(0.63139686,0.10067417,0.35664819);
  rgb(103pt)=(0.63812122,0.09938212,0.35603757);
  rgb(104pt)=(0.64483795,0.0980891,0.35533555);
  rgb(105pt)=(0.65154562,0.09680192,0.35454107);
  rgb(106pt)=(0.65824241,0.09552918,0.3536529);
  rgb(107pt)=(0.66492652,0.09428017,0.3526697);
  rgb(108pt)=(0.67159578,0.09306598,0.35159077);
  rgb(109pt)=(0.67824099,0.09192342,0.3504148);
  rgb(110pt)=(0.684863,0.09085633,0.34914061);
  rgb(111pt)=(0.69146268,0.0898675,0.34776864);
  rgb(112pt)=(0.69803757,0.08897226,0.3462986);
  rgb(113pt)=(0.70457834,0.0882129,0.34473046);
  rgb(114pt)=(0.71108138,0.08761223,0.3430635);
  rgb(115pt)=(0.7175507,0.08716212,0.34129974);
  rgb(116pt)=(0.72398193,0.08688725,0.33943958);
  rgb(117pt)=(0.73035829,0.0868623,0.33748452);
  rgb(118pt)=(0.73669146,0.08704683,0.33543669);
  rgb(119pt)=(0.74297501,0.08747196,0.33329799);
  rgb(120pt)=(0.74919318,0.08820542,0.33107204);
  rgb(121pt)=(0.75535825,0.08919792,0.32876184);
  rgb(122pt)=(0.76145589,0.09050716,0.32637117);
  rgb(123pt)=(0.76748424,0.09213602,0.32390525);
  rgb(124pt)=(0.77344838,0.09405684,0.32136808);
  rgb(125pt)=(0.77932641,0.09634794,0.31876642);
  rgb(126pt)=(0.78513609,0.09892473,0.31610488);
  rgb(127pt)=(0.79085854,0.10184672,0.313391);
  rgb(128pt)=(0.7965014,0.10506637,0.31063031);
  rgb(129pt)=(0.80205987,0.10858333,0.30783);
  rgb(130pt)=(0.80752799,0.11239964,0.30499738);
  rgb(131pt)=(0.81291606,0.11645784,0.30213802);
  rgb(132pt)=(0.81820481,0.12080606,0.29926105);
  rgb(133pt)=(0.82341472,0.12535343,0.2963705);
  rgb(134pt)=(0.82852822,0.13014118,0.29347474);
  rgb(135pt)=(0.83355779,0.13511035,0.29057852);
  rgb(136pt)=(0.83850183,0.14025098,0.2876878);
  rgb(137pt)=(0.84335441,0.14556683,0.28480819);
  rgb(138pt)=(0.84813096,0.15099892,0.281943);
  rgb(139pt)=(0.85281737,0.15657772,0.27909826);
  rgb(140pt)=(0.85742602,0.1622583,0.27627462);
  rgb(141pt)=(0.86196552,0.16801239,0.27346473);
  rgb(142pt)=(0.86641628,0.17387796,0.27070818);
  rgb(143pt)=(0.87079129,0.17982114,0.26797378);
  rgb(144pt)=(0.87507281,0.18587368,0.26529697);
  rgb(145pt)=(0.87925878,0.19203259,0.26268136);
  rgb(146pt)=(0.8833417,0.19830556,0.26014181);
  rgb(147pt)=(0.88731387,0.20469941,0.25769539);
  rgb(148pt)=(0.89116859,0.21121788,0.2553592);
  rgb(149pt)=(0.89490337,0.21785614,0.25314362);
  rgb(150pt)=(0.8985026,0.22463251,0.25108745);
  rgb(151pt)=(0.90197527,0.23152063,0.24918223);
  rgb(152pt)=(0.90530097,0.23854541,0.24748098);
  rgb(153pt)=(0.90848638,0.24568473,0.24598324);
  rgb(154pt)=(0.911533,0.25292623,0.24470258);
  rgb(155pt)=(0.9144225,0.26028902,0.24369359);
  rgb(156pt)=(0.91717106,0.26773821,0.24294137);
  rgb(157pt)=(0.91978131,0.27526191,0.24245973);
  rgb(158pt)=(0.92223947,0.28287251,0.24229568);
  rgb(159pt)=(0.92456587,0.29053388,0.24242622);
  rgb(160pt)=(0.92676657,0.29823282,0.24285536);
  rgb(161pt)=(0.92882964,0.30598085,0.24362274);
  rgb(162pt)=(0.93078135,0.31373977,0.24468803);
  rgb(163pt)=(0.93262051,0.3215093,0.24606461);
  rgb(164pt)=(0.93435067,0.32928362,0.24775328);
  rgb(165pt)=(0.93599076,0.33703942,0.24972157);
  rgb(166pt)=(0.93752831,0.34479177,0.25199928);
  rgb(167pt)=(0.93899289,0.35250734,0.25452808);
  rgb(168pt)=(0.94036561,0.36020899,0.25734661);
  rgb(169pt)=(0.94167588,0.36786594,0.2603949);
  rgb(170pt)=(0.94291042,0.37549479,0.26369821);
  rgb(171pt)=(0.94408513,0.3830811,0.26722004);
  rgb(172pt)=(0.94520419,0.39062329,0.27094924);
  rgb(173pt)=(0.94625977,0.39813168,0.27489742);
  rgb(174pt)=(0.94727016,0.4055909,0.27902322);
  rgb(175pt)=(0.94823505,0.41300424,0.28332283);
  rgb(176pt)=(0.94914549,0.42038251,0.28780969);
  rgb(177pt)=(0.95001704,0.42771398,0.29244728);
  rgb(178pt)=(0.95085121,0.43500005,0.29722817);
  rgb(179pt)=(0.95165009,0.44224144,0.30214494);
  rgb(180pt)=(0.9524044,0.44944853,0.3072105);
  rgb(181pt)=(0.95312556,0.45661389,0.31239776);
  rgb(182pt)=(0.95381595,0.46373781,0.31769923);
  rgb(183pt)=(0.95447591,0.47082238,0.32310953);
  rgb(184pt)=(0.95510255,0.47787236,0.32862553);
  rgb(185pt)=(0.95569679,0.48489115,0.33421404);
  rgb(186pt)=(0.95626788,0.49187351,0.33985601);
  rgb(187pt)=(0.95681685,0.49882008,0.34555431);
  rgb(188pt)=(0.9573439,0.50573243,0.35130912);
  rgb(189pt)=(0.95784842,0.51261283,0.35711942);
  rgb(190pt)=(0.95833051,0.51946267,0.36298589);
  rgb(191pt)=(0.95879054,0.52628305,0.36890904);
  rgb(192pt)=(0.95922872,0.53307513,0.3748895);
  rgb(193pt)=(0.95964538,0.53983991,0.38092784);
  rgb(194pt)=(0.96004345,0.54657593,0.3870292);
  rgb(195pt)=(0.96042097,0.55328624,0.39319057);
  rgb(196pt)=(0.96077819,0.55997184,0.39941173);
  rgb(197pt)=(0.9611152,0.5666337,0.40569343);
  rgb(198pt)=(0.96143273,0.57327231,0.41203603);
  rgb(199pt)=(0.96173392,0.57988594,0.41844491);
  rgb(200pt)=(0.96201757,0.58647675,0.42491751);
  rgb(201pt)=(0.96228344,0.59304598,0.43145271);
  rgb(202pt)=(0.96253168,0.5995944,0.43805131);
  rgb(203pt)=(0.96276513,0.60612062,0.44471698);
  rgb(204pt)=(0.96298491,0.6126247,0.45145074);
  rgb(205pt)=(0.96318967,0.61910879,0.45824902);
  rgb(206pt)=(0.96337949,0.6255736,0.46511271);
  rgb(207pt)=(0.96355923,0.63201624,0.47204746);
  rgb(208pt)=(0.96372785,0.63843852,0.47905028);
  rgb(209pt)=(0.96388426,0.64484214,0.4861196);
  rgb(210pt)=(0.96403203,0.65122535,0.4932578);
  rgb(211pt)=(0.96417332,0.65758729,0.50046894);
  rgb(212pt)=(0.9643063,0.66393045,0.5077467);
  rgb(213pt)=(0.96443322,0.67025402,0.51509334);
  rgb(214pt)=(0.96455845,0.67655564,0.52251447);
  rgb(215pt)=(0.96467922,0.68283846,0.53000231);
  rgb(216pt)=(0.96479861,0.68910113,0.53756026);
  rgb(217pt)=(0.96492035,0.69534192,0.5451917);
  rgb(218pt)=(0.96504223,0.7015636,0.5528892);
  rgb(219pt)=(0.96516917,0.70776351,0.5606593);
  rgb(220pt)=(0.96530224,0.71394212,0.56849894);
  rgb(221pt)=(0.96544032,0.72010124,0.57640375);
  rgb(222pt)=(0.96559206,0.72623592,0.58438387);
  rgb(223pt)=(0.96575293,0.73235058,0.59242739);
  rgb(224pt)=(0.96592829,0.73844258,0.60053991);
  rgb(225pt)=(0.96612013,0.74451182,0.60871954);
  rgb(226pt)=(0.96632832,0.75055966,0.61696136);
  rgb(227pt)=(0.96656022,0.75658231,0.62527295);
  rgb(228pt)=(0.96681185,0.76258381,0.63364277);
  rgb(229pt)=(0.96709183,0.76855969,0.64207921);
  rgb(230pt)=(0.96739773,0.77451297,0.65057302);
  rgb(231pt)=(0.96773482,0.78044149,0.65912731);
  rgb(232pt)=(0.96810471,0.78634563,0.66773889);
  rgb(233pt)=(0.96850919,0.79222565,0.6764046);
  rgb(234pt)=(0.96893132,0.79809112,0.68512266);
  rgb(235pt)=(0.96935926,0.80395415,0.69383201);
  rgb(236pt)=(0.9698028,0.80981139,0.70252255);
  rgb(237pt)=(0.97025511,0.81566605,0.71120296);
  rgb(238pt)=(0.97071849,0.82151775,0.71987163);
  rgb(239pt)=(0.97120159,0.82736371,0.72851999);
  rgb(240pt)=(0.97169389,0.83320847,0.73716071);
  rgb(241pt)=(0.97220061,0.83905052,0.74578903);
  rgb(242pt)=(0.97272597,0.84488881,0.75440141);
  rgb(243pt)=(0.97327085,0.85072354,0.76299805);
  rgb(244pt)=(0.97383206,0.85655639,0.77158353);
  rgb(245pt)=(0.97441222,0.86238689,0.78015619);
  rgb(246pt)=(0.97501782,0.86821321,0.78871034);
  rgb(247pt)=(0.97564391,0.87403763,0.79725261);
  rgb(248pt)=(0.97628674,0.87986189,0.8057883);
  rgb(249pt)=(0.97696114,0.88568129,0.81430324);
  rgb(250pt)=(0.97765722,0.89149971,0.82280948);
  rgb(251pt)=(0.97837585,0.89731727,0.83130786);
  rgb(252pt)=(0.97912374,0.90313207,0.83979337);
  rgb(253pt)=(0.979891,0.90894778,0.84827858);
  rgb(254pt)=(0.98067764,0.91476465,0.85676611);
  rgb(255pt)=(0.98137749,0.92061729,0.86536915)
},
point meta max=-0.896735214492083,
point meta min=-7.03741788026212,
tick align=outside,
tick pos=left,
x grid style={darkgray176},
xlabel={$\theta_1$},
xmin=0, xmax=40,
xtick style={color=black},
xtick={0,20,40},
xticklabels={$0$,$\pi$,$2\pi$},
y dir=reverse,
y grid style={darkgray176},
ylabel={$\theta_0$},
ymin=0, ymax=40,
ytick style={color=black},
ytick={0,20,40},
yticklabel style={rotate=90.0},
yticklabels={$2\pi$,$\pi$,$0$}
]
\addplot graphics[xmin=0, xmax=40, ymin=40, ymax=0] {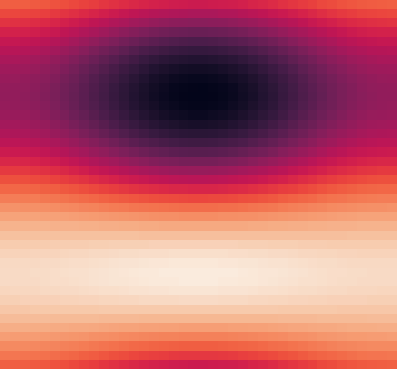};
\end{axis}

\end{tikzpicture}

%% file: tikz_figures/hobo_landscape.tex
% This file was created with tikzplotlib v0.10.1.
\begin{tikzpicture}

\definecolor{darkgray176}{RGB}{176,176,176}

\begin{axis}[
colorbar,
colorbar style={ylabel={Energy}, ylabel style={yshift=-0.1cm}},
colormap={mymap}{[1pt]
 rgb(0pt)=(0.01060815,0.01808215,0.10018654);
  rgb(1pt)=(0.01428972,0.02048237,0.10374486);
  rgb(2pt)=(0.01831941,0.0229766,0.10738511);
  rgb(3pt)=(0.02275049,0.02554464,0.11108639);
  rgb(4pt)=(0.02759119,0.02818316,0.11483751);
  rgb(5pt)=(0.03285175,0.03088792,0.11863035);
  rgb(6pt)=(0.03853466,0.03365771,0.12245873);
  rgb(7pt)=(0.04447016,0.03648425,0.12631831);
  rgb(8pt)=(0.05032105,0.03936808,0.13020508);
  rgb(9pt)=(0.05611171,0.04224835,0.13411624);
  rgb(10pt)=(0.0618531,0.04504866,0.13804929);
  rgb(11pt)=(0.06755457,0.04778179,0.14200206);
  rgb(12pt)=(0.0732236,0.05045047,0.14597263);
  rgb(13pt)=(0.0788708,0.05305461,0.14995981);
  rgb(14pt)=(0.08450105,0.05559631,0.15396203);
  rgb(15pt)=(0.09011319,0.05808059,0.15797687);
  rgb(16pt)=(0.09572396,0.06050127,0.16200507);
  rgb(17pt)=(0.10132312,0.06286782,0.16604287);
  rgb(18pt)=(0.10692823,0.06517224,0.17009175);
  rgb(19pt)=(0.1125315,0.06742194,0.17414848);
  rgb(20pt)=(0.11813947,0.06961499,0.17821272);
  rgb(21pt)=(0.12375803,0.07174938,0.18228425);
  rgb(22pt)=(0.12938228,0.07383015,0.18636053);
  rgb(23pt)=(0.13501631,0.07585609,0.19044109);
  rgb(24pt)=(0.14066867,0.0778224,0.19452676);
  rgb(25pt)=(0.14633406,0.07973393,0.1986151);
  rgb(26pt)=(0.15201338,0.08159108,0.20270523);
  rgb(27pt)=(0.15770877,0.08339312,0.20679668);
  rgb(28pt)=(0.16342174,0.0851396,0.21088893);
  rgb(29pt)=(0.16915387,0.08682996,0.21498104);
  rgb(30pt)=(0.17489524,0.08848235,0.2190294);
  rgb(31pt)=(0.18065495,0.09009031,0.22303512);
  rgb(32pt)=(0.18643324,0.09165431,0.22699705);
  rgb(33pt)=(0.19223028,0.09317479,0.23091409);
  rgb(34pt)=(0.19804623,0.09465217,0.23478512);
  rgb(35pt)=(0.20388117,0.09608689,0.23860907);
  rgb(36pt)=(0.20973515,0.09747934,0.24238489);
  rgb(37pt)=(0.21560818,0.09882993,0.24611154);
  rgb(38pt)=(0.22150014,0.10013944,0.2497868);
  rgb(39pt)=(0.22741085,0.10140876,0.25340813);
  rgb(40pt)=(0.23334047,0.10263737,0.25697736);
  rgb(41pt)=(0.23928891,0.10382562,0.2604936);
  rgb(42pt)=(0.24525608,0.10497384,0.26395596);
  rgb(43pt)=(0.25124182,0.10608236,0.26736359);
  rgb(44pt)=(0.25724602,0.10715148,0.27071569);
  rgb(45pt)=(0.26326851,0.1081815,0.27401148);
  rgb(46pt)=(0.26930915,0.1091727,0.2772502);
  rgb(47pt)=(0.27536766,0.11012568,0.28043021);
  rgb(48pt)=(0.28144375,0.11104133,0.2835489);
  rgb(49pt)=(0.2875374,0.11191896,0.28660853);
  rgb(50pt)=(0.29364846,0.11275876,0.2896085);
  rgb(51pt)=(0.29977678,0.11356089,0.29254823);
  rgb(52pt)=(0.30592213,0.11432553,0.29542718);
  rgb(53pt)=(0.31208435,0.11505284,0.29824485);
  rgb(54pt)=(0.31826327,0.1157429,0.30100076);
  rgb(55pt)=(0.32445869,0.11639585,0.30369448);
  rgb(56pt)=(0.33067031,0.11701189,0.30632563);
  rgb(57pt)=(0.33689808,0.11759095,0.3088938);
  rgb(58pt)=(0.34314168,0.11813362,0.31139721);
  rgb(59pt)=(0.34940101,0.11863987,0.3138355);
  rgb(60pt)=(0.355676,0.11910909,0.31620996);
  rgb(61pt)=(0.36196644,0.1195413,0.31852037);
  rgb(62pt)=(0.36827206,0.11993653,0.32076656);
  rgb(63pt)=(0.37459292,0.12029443,0.32294825);
  rgb(64pt)=(0.38092887,0.12061482,0.32506528);
  rgb(65pt)=(0.38727975,0.12089756,0.3271175);
  rgb(66pt)=(0.39364518,0.12114272,0.32910494);
  rgb(67pt)=(0.40002537,0.12134964,0.33102734);
  rgb(68pt)=(0.40642019,0.12151801,0.33288464);
  rgb(69pt)=(0.41282936,0.12164769,0.33467689);
  rgb(70pt)=(0.41925278,0.12173833,0.33640407);
  rgb(71pt)=(0.42569057,0.12178916,0.33806605);
  rgb(72pt)=(0.43214263,0.12179973,0.33966284);
  rgb(73pt)=(0.43860848,0.12177004,0.34119475);
  rgb(74pt)=(0.44508855,0.12169883,0.34266151);
  rgb(75pt)=(0.45158266,0.12158557,0.34406324);
  rgb(76pt)=(0.45809049,0.12142996,0.34540024);
  rgb(77pt)=(0.46461238,0.12123063,0.34667231);
  rgb(78pt)=(0.47114798,0.12098721,0.34787978);
  rgb(79pt)=(0.47769736,0.12069864,0.34902273);
  rgb(80pt)=(0.48426077,0.12036349,0.35010104);
  rgb(81pt)=(0.49083761,0.11998161,0.35111537);
  rgb(82pt)=(0.49742847,0.11955087,0.35206533);
  rgb(83pt)=(0.50403286,0.11907081,0.35295152);
  rgb(84pt)=(0.51065109,0.11853959,0.35377385);
  rgb(85pt)=(0.51728314,0.1179558,0.35453252);
  rgb(86pt)=(0.52392883,0.11731817,0.35522789);
  rgb(87pt)=(0.53058853,0.11662445,0.35585982);
  rgb(88pt)=(0.53726173,0.11587369,0.35642903);
  rgb(89pt)=(0.54394898,0.11506307,0.35693521);
  rgb(90pt)=(0.5506426,0.11420757,0.35737863);
  rgb(91pt)=(0.55734473,0.11330456,0.35775059);
  rgb(92pt)=(0.56405586,0.11235265,0.35804813);
  rgb(93pt)=(0.57077365,0.11135597,0.35827146);
  rgb(94pt)=(0.5774991,0.11031233,0.35841679);
  rgb(95pt)=(0.58422945,0.10922707,0.35848469);
  rgb(96pt)=(0.59096382,0.10810205,0.35847347);
  rgb(97pt)=(0.59770215,0.10693774,0.35838029);
  rgb(98pt)=(0.60444226,0.10573912,0.35820487);
  rgb(99pt)=(0.61118304,0.10450943,0.35794557);
  rgb(100pt)=(0.61792306,0.10325288,0.35760108);
  rgb(101pt)=(0.62466162,0.10197244,0.35716891);
  rgb(102pt)=(0.63139686,0.10067417,0.35664819);
  rgb(103pt)=(0.63812122,0.09938212,0.35603757);
  rgb(104pt)=(0.64483795,0.0980891,0.35533555);
  rgb(105pt)=(0.65154562,0.09680192,0.35454107);
  rgb(106pt)=(0.65824241,0.09552918,0.3536529);
  rgb(107pt)=(0.66492652,0.09428017,0.3526697);
  rgb(108pt)=(0.67159578,0.09306598,0.35159077);
  rgb(109pt)=(0.67824099,0.09192342,0.3504148);
  rgb(110pt)=(0.684863,0.09085633,0.34914061);
  rgb(111pt)=(0.69146268,0.0898675,0.34776864);
  rgb(112pt)=(0.69803757,0.08897226,0.3462986);
  rgb(113pt)=(0.70457834,0.0882129,0.34473046);
  rgb(114pt)=(0.71108138,0.08761223,0.3430635);
  rgb(115pt)=(0.7175507,0.08716212,0.34129974);
  rgb(116pt)=(0.72398193,0.08688725,0.33943958);
  rgb(117pt)=(0.73035829,0.0868623,0.33748452);
  rgb(118pt)=(0.73669146,0.08704683,0.33543669);
  rgb(119pt)=(0.74297501,0.08747196,0.33329799);
  rgb(120pt)=(0.74919318,0.08820542,0.33107204);
  rgb(121pt)=(0.75535825,0.08919792,0.32876184);
  rgb(122pt)=(0.76145589,0.09050716,0.32637117);
  rgb(123pt)=(0.76748424,0.09213602,0.32390525);
  rgb(124pt)=(0.77344838,0.09405684,0.32136808);
  rgb(125pt)=(0.77932641,0.09634794,0.31876642);
  rgb(126pt)=(0.78513609,0.09892473,0.31610488);
  rgb(127pt)=(0.79085854,0.10184672,0.313391);
  rgb(128pt)=(0.7965014,0.10506637,0.31063031);
  rgb(129pt)=(0.80205987,0.10858333,0.30783);
  rgb(130pt)=(0.80752799,0.11239964,0.30499738);
  rgb(131pt)=(0.81291606,0.11645784,0.30213802);
  rgb(132pt)=(0.81820481,0.12080606,0.29926105);
  rgb(133pt)=(0.82341472,0.12535343,0.2963705);
  rgb(134pt)=(0.82852822,0.13014118,0.29347474);
  rgb(135pt)=(0.83355779,0.13511035,0.29057852);
  rgb(136pt)=(0.83850183,0.14025098,0.2876878);
  rgb(137pt)=(0.84335441,0.14556683,0.28480819);
  rgb(138pt)=(0.84813096,0.15099892,0.281943);
  rgb(139pt)=(0.85281737,0.15657772,0.27909826);
  rgb(140pt)=(0.85742602,0.1622583,0.27627462);
  rgb(141pt)=(0.86196552,0.16801239,0.27346473);
  rgb(142pt)=(0.86641628,0.17387796,0.27070818);
  rgb(143pt)=(0.87079129,0.17982114,0.26797378);
  rgb(144pt)=(0.87507281,0.18587368,0.26529697);
  rgb(145pt)=(0.87925878,0.19203259,0.26268136);
  rgb(146pt)=(0.8833417,0.19830556,0.26014181);
  rgb(147pt)=(0.88731387,0.20469941,0.25769539);
  rgb(148pt)=(0.89116859,0.21121788,0.2553592);
  rgb(149pt)=(0.89490337,0.21785614,0.25314362);
  rgb(150pt)=(0.8985026,0.22463251,0.25108745);
  rgb(151pt)=(0.90197527,0.23152063,0.24918223);
  rgb(152pt)=(0.90530097,0.23854541,0.24748098);
  rgb(153pt)=(0.90848638,0.24568473,0.24598324);
  rgb(154pt)=(0.911533,0.25292623,0.24470258);
  rgb(155pt)=(0.9144225,0.26028902,0.24369359);
  rgb(156pt)=(0.91717106,0.26773821,0.24294137);
  rgb(157pt)=(0.91978131,0.27526191,0.24245973);
  rgb(158pt)=(0.92223947,0.28287251,0.24229568);
  rgb(159pt)=(0.92456587,0.29053388,0.24242622);
  rgb(160pt)=(0.92676657,0.29823282,0.24285536);
  rgb(161pt)=(0.92882964,0.30598085,0.24362274);
  rgb(162pt)=(0.93078135,0.31373977,0.24468803);
  rgb(163pt)=(0.93262051,0.3215093,0.24606461);
  rgb(164pt)=(0.93435067,0.32928362,0.24775328);
  rgb(165pt)=(0.93599076,0.33703942,0.24972157);
  rgb(166pt)=(0.93752831,0.34479177,0.25199928);
  rgb(167pt)=(0.93899289,0.35250734,0.25452808);
  rgb(168pt)=(0.94036561,0.36020899,0.25734661);
  rgb(169pt)=(0.94167588,0.36786594,0.2603949);
  rgb(170pt)=(0.94291042,0.37549479,0.26369821);
  rgb(171pt)=(0.94408513,0.3830811,0.26722004);
  rgb(172pt)=(0.94520419,0.39062329,0.27094924);
  rgb(173pt)=(0.94625977,0.39813168,0.27489742);
  rgb(174pt)=(0.94727016,0.4055909,0.27902322);
  rgb(175pt)=(0.94823505,0.41300424,0.28332283);
  rgb(176pt)=(0.94914549,0.42038251,0.28780969);
  rgb(177pt)=(0.95001704,0.42771398,0.29244728);
  rgb(178pt)=(0.95085121,0.43500005,0.29722817);
  rgb(179pt)=(0.95165009,0.44224144,0.30214494);
  rgb(180pt)=(0.9524044,0.44944853,0.3072105);
  rgb(181pt)=(0.95312556,0.45661389,0.31239776);
  rgb(182pt)=(0.95381595,0.46373781,0.31769923);
  rgb(183pt)=(0.95447591,0.47082238,0.32310953);
  rgb(184pt)=(0.95510255,0.47787236,0.32862553);
  rgb(185pt)=(0.95569679,0.48489115,0.33421404);
  rgb(186pt)=(0.95626788,0.49187351,0.33985601);
  rgb(187pt)=(0.95681685,0.49882008,0.34555431);
  rgb(188pt)=(0.9573439,0.50573243,0.35130912);
  rgb(189pt)=(0.95784842,0.51261283,0.35711942);
  rgb(190pt)=(0.95833051,0.51946267,0.36298589);
  rgb(191pt)=(0.95879054,0.52628305,0.36890904);
  rgb(192pt)=(0.95922872,0.53307513,0.3748895);
  rgb(193pt)=(0.95964538,0.53983991,0.38092784);
  rgb(194pt)=(0.96004345,0.54657593,0.3870292);
  rgb(195pt)=(0.96042097,0.55328624,0.39319057);
  rgb(196pt)=(0.96077819,0.55997184,0.39941173);
  rgb(197pt)=(0.9611152,0.5666337,0.40569343);
  rgb(198pt)=(0.96143273,0.57327231,0.41203603);
  rgb(199pt)=(0.96173392,0.57988594,0.41844491);
  rgb(200pt)=(0.96201757,0.58647675,0.42491751);
  rgb(201pt)=(0.96228344,0.59304598,0.43145271);
  rgb(202pt)=(0.96253168,0.5995944,0.43805131);
  rgb(203pt)=(0.96276513,0.60612062,0.44471698);
  rgb(204pt)=(0.96298491,0.6126247,0.45145074);
  rgb(205pt)=(0.96318967,0.61910879,0.45824902);
  rgb(206pt)=(0.96337949,0.6255736,0.46511271);
  rgb(207pt)=(0.96355923,0.63201624,0.47204746);
  rgb(208pt)=(0.96372785,0.63843852,0.47905028);
  rgb(209pt)=(0.96388426,0.64484214,0.4861196);
  rgb(210pt)=(0.96403203,0.65122535,0.4932578);
  rgb(211pt)=(0.96417332,0.65758729,0.50046894);
  rgb(212pt)=(0.9643063,0.66393045,0.5077467);
  rgb(213pt)=(0.96443322,0.67025402,0.51509334);
  rgb(214pt)=(0.96455845,0.67655564,0.52251447);
  rgb(215pt)=(0.96467922,0.68283846,0.53000231);
  rgb(216pt)=(0.96479861,0.68910113,0.53756026);
  rgb(217pt)=(0.96492035,0.69534192,0.5451917);
  rgb(218pt)=(0.96504223,0.7015636,0.5528892);
  rgb(219pt)=(0.96516917,0.70776351,0.5606593);
  rgb(220pt)=(0.96530224,0.71394212,0.56849894);
  rgb(221pt)=(0.96544032,0.72010124,0.57640375);
  rgb(222pt)=(0.96559206,0.72623592,0.58438387);
  rgb(223pt)=(0.96575293,0.73235058,0.59242739);
  rgb(224pt)=(0.96592829,0.73844258,0.60053991);
  rgb(225pt)=(0.96612013,0.74451182,0.60871954);
  rgb(226pt)=(0.96632832,0.75055966,0.61696136);
  rgb(227pt)=(0.96656022,0.75658231,0.62527295);
  rgb(228pt)=(0.96681185,0.76258381,0.63364277);
  rgb(229pt)=(0.96709183,0.76855969,0.64207921);
  rgb(230pt)=(0.96739773,0.77451297,0.65057302);
  rgb(231pt)=(0.96773482,0.78044149,0.65912731);
  rgb(232pt)=(0.96810471,0.78634563,0.66773889);
  rgb(233pt)=(0.96850919,0.79222565,0.6764046);
  rgb(234pt)=(0.96893132,0.79809112,0.68512266);
  rgb(235pt)=(0.96935926,0.80395415,0.69383201);
  rgb(236pt)=(0.9698028,0.80981139,0.70252255);
  rgb(237pt)=(0.97025511,0.81566605,0.71120296);
  rgb(238pt)=(0.97071849,0.82151775,0.71987163);
  rgb(239pt)=(0.97120159,0.82736371,0.72851999);
  rgb(240pt)=(0.97169389,0.83320847,0.73716071);
  rgb(241pt)=(0.97220061,0.83905052,0.74578903);
  rgb(242pt)=(0.97272597,0.84488881,0.75440141);
  rgb(243pt)=(0.97327085,0.85072354,0.76299805);
  rgb(244pt)=(0.97383206,0.85655639,0.77158353);
  rgb(245pt)=(0.97441222,0.86238689,0.78015619);
  rgb(246pt)=(0.97501782,0.86821321,0.78871034);
  rgb(247pt)=(0.97564391,0.87403763,0.79725261);
  rgb(248pt)=(0.97628674,0.87986189,0.8057883);
  rgb(249pt)=(0.97696114,0.88568129,0.81430324);
  rgb(250pt)=(0.97765722,0.89149971,0.82280948);
  rgb(251pt)=(0.97837585,0.89731727,0.83130786);
  rgb(252pt)=(0.97912374,0.90313207,0.83979337);
  rgb(253pt)=(0.979891,0.90894778,0.84827858);
  rgb(254pt)=(0.98067764,0.91476465,0.85676611);
  rgb(255pt)=(0.98137749,0.92061729,0.86536915)
},
point meta max=2.47367597073077,
point meta min=1.68164769113216,
tick align=outside,
tick pos=left,
x grid style={darkgray176},
xlabel={$\theta_1$},
xmin=0, xmax=40,
xtick style={color=black},
xtick={0,20,40},
xticklabels={$0$,$\pi$,$2\pi$},
y dir=reverse,
y grid style={darkgray176},
ylabel={$\theta_0$},
ymin=0, ymax=40,
ytick style={color=black},
ytick={0,20,40},
yticklabel style={rotate=90.0},
yticklabels={$2\pi$,$\pi$,$0$}
]
\addplot graphics[xmin=0, xmax=40, ymin=40, ymax=0] {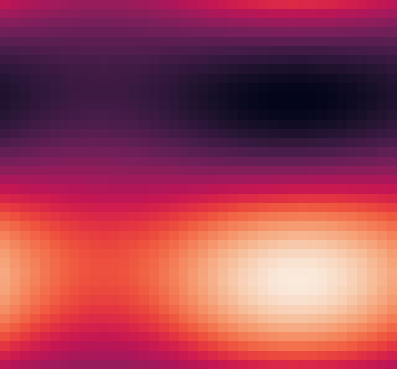};
\end{axis}

\end{tikzpicture}

%% file: tikz_figures/permutation_landscape.tex
% This file was created with tikzplotlib v0.10.1.
\begin{tikzpicture}

\definecolor{darkgray176}{RGB}{176,176,176}

\begin{axis}[
colorbar,
colorbar style={ylabel={Energy}, ylabel style={yshift=-0.1cm}},
colormap={mymap}{[1pt]
 rgb(0pt)=(0.01060815,0.01808215,0.10018654);
  rgb(1pt)=(0.01428972,0.02048237,0.10374486);
  rgb(2pt)=(0.01831941,0.0229766,0.10738511);
  rgb(3pt)=(0.02275049,0.02554464,0.11108639);
  rgb(4pt)=(0.02759119,0.02818316,0.11483751);
  rgb(5pt)=(0.03285175,0.03088792,0.11863035);
  rgb(6pt)=(0.03853466,0.03365771,0.12245873);
  rgb(7pt)=(0.04447016,0.03648425,0.12631831);
  rgb(8pt)=(0.05032105,0.03936808,0.13020508);
  rgb(9pt)=(0.05611171,0.04224835,0.13411624);
  rgb(10pt)=(0.0618531,0.04504866,0.13804929);
  rgb(11pt)=(0.06755457,0.04778179,0.14200206);
  rgb(12pt)=(0.0732236,0.05045047,0.14597263);
  rgb(13pt)=(0.0788708,0.05305461,0.14995981);
  rgb(14pt)=(0.08450105,0.05559631,0.15396203);
  rgb(15pt)=(0.09011319,0.05808059,0.15797687);
  rgb(16pt)=(0.09572396,0.06050127,0.16200507);
  rgb(17pt)=(0.10132312,0.06286782,0.16604287);
  rgb(18pt)=(0.10692823,0.06517224,0.17009175);
  rgb(19pt)=(0.1125315,0.06742194,0.17414848);
  rgb(20pt)=(0.11813947,0.06961499,0.17821272);
  rgb(21pt)=(0.12375803,0.07174938,0.18228425);
  rgb(22pt)=(0.12938228,0.07383015,0.18636053);
  rgb(23pt)=(0.13501631,0.07585609,0.19044109);
  rgb(24pt)=(0.14066867,0.0778224,0.19452676);
  rgb(25pt)=(0.14633406,0.07973393,0.1986151);
  rgb(26pt)=(0.15201338,0.08159108,0.20270523);
  rgb(27pt)=(0.15770877,0.08339312,0.20679668);
  rgb(28pt)=(0.16342174,0.0851396,0.21088893);
  rgb(29pt)=(0.16915387,0.08682996,0.21498104);
  rgb(30pt)=(0.17489524,0.08848235,0.2190294);
  rgb(31pt)=(0.18065495,0.09009031,0.22303512);
  rgb(32pt)=(0.18643324,0.09165431,0.22699705);
  rgb(33pt)=(0.19223028,0.09317479,0.23091409);
  rgb(34pt)=(0.19804623,0.09465217,0.23478512);
  rgb(35pt)=(0.20388117,0.09608689,0.23860907);
  rgb(36pt)=(0.20973515,0.09747934,0.24238489);
  rgb(37pt)=(0.21560818,0.09882993,0.24611154);
  rgb(38pt)=(0.22150014,0.10013944,0.2497868);
  rgb(39pt)=(0.22741085,0.10140876,0.25340813);
  rgb(40pt)=(0.23334047,0.10263737,0.25697736);
  rgb(41pt)=(0.23928891,0.10382562,0.2604936);
  rgb(42pt)=(0.24525608,0.10497384,0.26395596);
  rgb(43pt)=(0.25124182,0.10608236,0.26736359);
  rgb(44pt)=(0.25724602,0.10715148,0.27071569);
  rgb(45pt)=(0.26326851,0.1081815,0.27401148);
  rgb(46pt)=(0.26930915,0.1091727,0.2772502);
  rgb(47pt)=(0.27536766,0.11012568,0.28043021);
  rgb(48pt)=(0.28144375,0.11104133,0.2835489);
  rgb(49pt)=(0.2875374,0.11191896,0.28660853);
  rgb(50pt)=(0.29364846,0.11275876,0.2896085);
  rgb(51pt)=(0.29977678,0.11356089,0.29254823);
  rgb(52pt)=(0.30592213,0.11432553,0.29542718);
  rgb(53pt)=(0.31208435,0.11505284,0.29824485);
  rgb(54pt)=(0.31826327,0.1157429,0.30100076);
  rgb(55pt)=(0.32445869,0.11639585,0.30369448);
  rgb(56pt)=(0.33067031,0.11701189,0.30632563);
  rgb(57pt)=(0.33689808,0.11759095,0.3088938);
  rgb(58pt)=(0.34314168,0.11813362,0.31139721);
  rgb(59pt)=(0.34940101,0.11863987,0.3138355);
  rgb(60pt)=(0.355676,0.11910909,0.31620996);
  rgb(61pt)=(0.36196644,0.1195413,0.31852037);
  rgb(62pt)=(0.36827206,0.11993653,0.32076656);
  rgb(63pt)=(0.37459292,0.12029443,0.32294825);
  rgb(64pt)=(0.38092887,0.12061482,0.32506528);
  rgb(65pt)=(0.38727975,0.12089756,0.3271175);
  rgb(66pt)=(0.39364518,0.12114272,0.32910494);
  rgb(67pt)=(0.40002537,0.12134964,0.33102734);
  rgb(68pt)=(0.40642019,0.12151801,0.33288464);
  rgb(69pt)=(0.41282936,0.12164769,0.33467689);
  rgb(70pt)=(0.41925278,0.12173833,0.33640407);
  rgb(71pt)=(0.42569057,0.12178916,0.33806605);
  rgb(72pt)=(0.43214263,0.12179973,0.33966284);
  rgb(73pt)=(0.43860848,0.12177004,0.34119475);
  rgb(74pt)=(0.44508855,0.12169883,0.34266151);
  rgb(75pt)=(0.45158266,0.12158557,0.34406324);
  rgb(76pt)=(0.45809049,0.12142996,0.34540024);
  rgb(77pt)=(0.46461238,0.12123063,0.34667231);
  rgb(78pt)=(0.47114798,0.12098721,0.34787978);
  rgb(79pt)=(0.47769736,0.12069864,0.34902273);
  rgb(80pt)=(0.48426077,0.12036349,0.35010104);
  rgb(81pt)=(0.49083761,0.11998161,0.35111537);
  rgb(82pt)=(0.49742847,0.11955087,0.35206533);
  rgb(83pt)=(0.50403286,0.11907081,0.35295152);
  rgb(84pt)=(0.51065109,0.11853959,0.35377385);
  rgb(85pt)=(0.51728314,0.1179558,0.35453252);
  rgb(86pt)=(0.52392883,0.11731817,0.35522789);
  rgb(87pt)=(0.53058853,0.11662445,0.35585982);
  rgb(88pt)=(0.53726173,0.11587369,0.35642903);
  rgb(89pt)=(0.54394898,0.11506307,0.35693521);
  rgb(90pt)=(0.5506426,0.11420757,0.35737863);
  rgb(91pt)=(0.55734473,0.11330456,0.35775059);
  rgb(92pt)=(0.56405586,0.11235265,0.35804813);
  rgb(93pt)=(0.57077365,0.11135597,0.35827146);
  rgb(94pt)=(0.5774991,0.11031233,0.35841679);
  rgb(95pt)=(0.58422945,0.10922707,0.35848469);
  rgb(96pt)=(0.59096382,0.10810205,0.35847347);
  rgb(97pt)=(0.59770215,0.10693774,0.35838029);
  rgb(98pt)=(0.60444226,0.10573912,0.35820487);
  rgb(99pt)=(0.61118304,0.10450943,0.35794557);
  rgb(100pt)=(0.61792306,0.10325288,0.35760108);
  rgb(101pt)=(0.62466162,0.10197244,0.35716891);
  rgb(102pt)=(0.63139686,0.10067417,0.35664819);
  rgb(103pt)=(0.63812122,0.09938212,0.35603757);
  rgb(104pt)=(0.64483795,0.0980891,0.35533555);
  rgb(105pt)=(0.65154562,0.09680192,0.35454107);
  rgb(106pt)=(0.65824241,0.09552918,0.3536529);
  rgb(107pt)=(0.66492652,0.09428017,0.3526697);
  rgb(108pt)=(0.67159578,0.09306598,0.35159077);
  rgb(109pt)=(0.67824099,0.09192342,0.3504148);
  rgb(110pt)=(0.684863,0.09085633,0.34914061);
  rgb(111pt)=(0.69146268,0.0898675,0.34776864);
  rgb(112pt)=(0.69803757,0.08897226,0.3462986);
  rgb(113pt)=(0.70457834,0.0882129,0.34473046);
  rgb(114pt)=(0.71108138,0.08761223,0.3430635);
  rgb(115pt)=(0.7175507,0.08716212,0.34129974);
  rgb(116pt)=(0.72398193,0.08688725,0.33943958);
  rgb(117pt)=(0.73035829,0.0868623,0.33748452);
  rgb(118pt)=(0.73669146,0.08704683,0.33543669);
  rgb(119pt)=(0.74297501,0.08747196,0.33329799);
  rgb(120pt)=(0.74919318,0.08820542,0.33107204);
  rgb(121pt)=(0.75535825,0.08919792,0.32876184);
  rgb(122pt)=(0.76145589,0.09050716,0.32637117);
  rgb(123pt)=(0.76748424,0.09213602,0.32390525);
  rgb(124pt)=(0.77344838,0.09405684,0.32136808);
  rgb(125pt)=(0.77932641,0.09634794,0.31876642);
  rgb(126pt)=(0.78513609,0.09892473,0.31610488);
  rgb(127pt)=(0.79085854,0.10184672,0.313391);
  rgb(128pt)=(0.7965014,0.10506637,0.31063031);
  rgb(129pt)=(0.80205987,0.10858333,0.30783);
  rgb(130pt)=(0.80752799,0.11239964,0.30499738);
  rgb(131pt)=(0.81291606,0.11645784,0.30213802);
  rgb(132pt)=(0.81820481,0.12080606,0.29926105);
  rgb(133pt)=(0.82341472,0.12535343,0.2963705);
  rgb(134pt)=(0.82852822,0.13014118,0.29347474);
  rgb(135pt)=(0.83355779,0.13511035,0.29057852);
  rgb(136pt)=(0.83850183,0.14025098,0.2876878);
  rgb(137pt)=(0.84335441,0.14556683,0.28480819);
  rgb(138pt)=(0.84813096,0.15099892,0.281943);
  rgb(139pt)=(0.85281737,0.15657772,0.27909826);
  rgb(140pt)=(0.85742602,0.1622583,0.27627462);
  rgb(141pt)=(0.86196552,0.16801239,0.27346473);
  rgb(142pt)=(0.86641628,0.17387796,0.27070818);
  rgb(143pt)=(0.87079129,0.17982114,0.26797378);
  rgb(144pt)=(0.87507281,0.18587368,0.26529697);
  rgb(145pt)=(0.87925878,0.19203259,0.26268136);
  rgb(146pt)=(0.8833417,0.19830556,0.26014181);
  rgb(147pt)=(0.88731387,0.20469941,0.25769539);
  rgb(148pt)=(0.89116859,0.21121788,0.2553592);
  rgb(149pt)=(0.89490337,0.21785614,0.25314362);
  rgb(150pt)=(0.8985026,0.22463251,0.25108745);
  rgb(151pt)=(0.90197527,0.23152063,0.24918223);
  rgb(152pt)=(0.90530097,0.23854541,0.24748098);
  rgb(153pt)=(0.90848638,0.24568473,0.24598324);
  rgb(154pt)=(0.911533,0.25292623,0.24470258);
  rgb(155pt)=(0.9144225,0.26028902,0.24369359);
  rgb(156pt)=(0.91717106,0.26773821,0.24294137);
  rgb(157pt)=(0.91978131,0.27526191,0.24245973);
  rgb(158pt)=(0.92223947,0.28287251,0.24229568);
  rgb(159pt)=(0.92456587,0.29053388,0.24242622);
  rgb(160pt)=(0.92676657,0.29823282,0.24285536);
  rgb(161pt)=(0.92882964,0.30598085,0.24362274);
  rgb(162pt)=(0.93078135,0.31373977,0.24468803);
  rgb(163pt)=(0.93262051,0.3215093,0.24606461);
  rgb(164pt)=(0.93435067,0.32928362,0.24775328);
  rgb(165pt)=(0.93599076,0.33703942,0.24972157);
  rgb(166pt)=(0.93752831,0.34479177,0.25199928);
  rgb(167pt)=(0.93899289,0.35250734,0.25452808);
  rgb(168pt)=(0.94036561,0.36020899,0.25734661);
  rgb(169pt)=(0.94167588,0.36786594,0.2603949);
  rgb(170pt)=(0.94291042,0.37549479,0.26369821);
  rgb(171pt)=(0.94408513,0.3830811,0.26722004);
  rgb(172pt)=(0.94520419,0.39062329,0.27094924);
  rgb(173pt)=(0.94625977,0.39813168,0.27489742);
  rgb(174pt)=(0.94727016,0.4055909,0.27902322);
  rgb(175pt)=(0.94823505,0.41300424,0.28332283);
  rgb(176pt)=(0.94914549,0.42038251,0.28780969);
  rgb(177pt)=(0.95001704,0.42771398,0.29244728);
  rgb(178pt)=(0.95085121,0.43500005,0.29722817);
  rgb(179pt)=(0.95165009,0.44224144,0.30214494);
  rgb(180pt)=(0.9524044,0.44944853,0.3072105);
  rgb(181pt)=(0.95312556,0.45661389,0.31239776);
  rgb(182pt)=(0.95381595,0.46373781,0.31769923);
  rgb(183pt)=(0.95447591,0.47082238,0.32310953);
  rgb(184pt)=(0.95510255,0.47787236,0.32862553);
  rgb(185pt)=(0.95569679,0.48489115,0.33421404);
  rgb(186pt)=(0.95626788,0.49187351,0.33985601);
  rgb(187pt)=(0.95681685,0.49882008,0.34555431);
  rgb(188pt)=(0.9573439,0.50573243,0.35130912);
  rgb(189pt)=(0.95784842,0.51261283,0.35711942);
  rgb(190pt)=(0.95833051,0.51946267,0.36298589);
  rgb(191pt)=(0.95879054,0.52628305,0.36890904);
  rgb(192pt)=(0.95922872,0.53307513,0.3748895);
  rgb(193pt)=(0.95964538,0.53983991,0.38092784);
  rgb(194pt)=(0.96004345,0.54657593,0.3870292);
  rgb(195pt)=(0.96042097,0.55328624,0.39319057);
  rgb(196pt)=(0.96077819,0.55997184,0.39941173);
  rgb(197pt)=(0.9611152,0.5666337,0.40569343);
  rgb(198pt)=(0.96143273,0.57327231,0.41203603);
  rgb(199pt)=(0.96173392,0.57988594,0.41844491);
  rgb(200pt)=(0.96201757,0.58647675,0.42491751);
  rgb(201pt)=(0.96228344,0.59304598,0.43145271);
  rgb(202pt)=(0.96253168,0.5995944,0.43805131);
  rgb(203pt)=(0.96276513,0.60612062,0.44471698);
  rgb(204pt)=(0.96298491,0.6126247,0.45145074);
  rgb(205pt)=(0.96318967,0.61910879,0.45824902);
  rgb(206pt)=(0.96337949,0.6255736,0.46511271);
  rgb(207pt)=(0.96355923,0.63201624,0.47204746);
  rgb(208pt)=(0.96372785,0.63843852,0.47905028);
  rgb(209pt)=(0.96388426,0.64484214,0.4861196);
  rgb(210pt)=(0.96403203,0.65122535,0.4932578);
  rgb(211pt)=(0.96417332,0.65758729,0.50046894);
  rgb(212pt)=(0.9643063,0.66393045,0.5077467);
  rgb(213pt)=(0.96443322,0.67025402,0.51509334);
  rgb(214pt)=(0.96455845,0.67655564,0.52251447);
  rgb(215pt)=(0.96467922,0.68283846,0.53000231);
  rgb(216pt)=(0.96479861,0.68910113,0.53756026);
  rgb(217pt)=(0.96492035,0.69534192,0.5451917);
  rgb(218pt)=(0.96504223,0.7015636,0.5528892);
  rgb(219pt)=(0.96516917,0.70776351,0.5606593);
  rgb(220pt)=(0.96530224,0.71394212,0.56849894);
  rgb(221pt)=(0.96544032,0.72010124,0.57640375);
  rgb(222pt)=(0.96559206,0.72623592,0.58438387);
  rgb(223pt)=(0.96575293,0.73235058,0.59242739);
  rgb(224pt)=(0.96592829,0.73844258,0.60053991);
  rgb(225pt)=(0.96612013,0.74451182,0.60871954);
  rgb(226pt)=(0.96632832,0.75055966,0.61696136);
  rgb(227pt)=(0.96656022,0.75658231,0.62527295);
  rgb(228pt)=(0.96681185,0.76258381,0.63364277);
  rgb(229pt)=(0.96709183,0.76855969,0.64207921);
  rgb(230pt)=(0.96739773,0.77451297,0.65057302);
  rgb(231pt)=(0.96773482,0.78044149,0.65912731);
  rgb(232pt)=(0.96810471,0.78634563,0.66773889);
  rgb(233pt)=(0.96850919,0.79222565,0.6764046);
  rgb(234pt)=(0.96893132,0.79809112,0.68512266);
  rgb(235pt)=(0.96935926,0.80395415,0.69383201);
  rgb(236pt)=(0.9698028,0.80981139,0.70252255);
  rgb(237pt)=(0.97025511,0.81566605,0.71120296);
  rgb(238pt)=(0.97071849,0.82151775,0.71987163);
  rgb(239pt)=(0.97120159,0.82736371,0.72851999);
  rgb(240pt)=(0.97169389,0.83320847,0.73716071);
  rgb(241pt)=(0.97220061,0.83905052,0.74578903);
  rgb(242pt)=(0.97272597,0.84488881,0.75440141);
  rgb(243pt)=(0.97327085,0.85072354,0.76299805);
  rgb(244pt)=(0.97383206,0.85655639,0.77158353);
  rgb(245pt)=(0.97441222,0.86238689,0.78015619);
  rgb(246pt)=(0.97501782,0.86821321,0.78871034);
  rgb(247pt)=(0.97564391,0.87403763,0.79725261);
  rgb(248pt)=(0.97628674,0.87986189,0.8057883);
  rgb(249pt)=(0.97696114,0.88568129,0.81430324);
  rgb(250pt)=(0.97765722,0.89149971,0.82280948);
  rgb(251pt)=(0.97837585,0.89731727,0.83130786);
  rgb(252pt)=(0.97912374,0.90313207,0.83979337);
  rgb(253pt)=(0.979891,0.90894778,0.84827858);
  rgb(254pt)=(0.98067764,0.91476465,0.85676611);
  rgb(255pt)=(0.98137749,0.92061729,0.86536915)
},
point meta max=180.347885469076,
point meta min=153.004639539386,
tick align=outside,
tick pos=left,
x grid style={darkgray176},
xlabel={$\theta_1$},
xmin=0, xmax=40,
xtick style={color=black},
xtick={0,20,40},
xticklabels={$0$,$\pi$,$2\pi$},
y dir=reverse,
y grid style={darkgray176},
ylabel={$\theta_0$},
ymin=0, ymax=40,
ytick style={color=black},
ytick={0,20,40},
yticklabel style={rotate=90.0},
yticklabels={$2\pi$,$\pi$,$0$}
]
\addplot graphics[xmin=0, xmax=40, ymin=40, ymax=0] {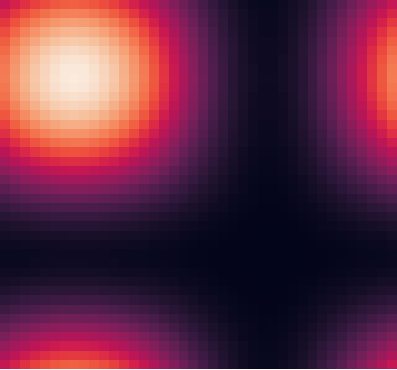};
\end{axis}

\end{tikzpicture}

%% file: tikz_figures/qubo_qaoa_landscape.tex
% This file was created with tikzplotlib v0.10.1.
\begin{tikzpicture}

\definecolor{darkgray176}{RGB}{176,176,176}

\begin{axis}[
colorbar,
colorbar style={ylabel={Energy}, ylabel style={yshift=0cm}},
colormap={mymap}{[1pt]
 rgb(0pt)=(0.01060815,0.01808215,0.10018654);
  rgb(1pt)=(0.01428972,0.02048237,0.10374486);
  rgb(2pt)=(0.01831941,0.0229766,0.10738511);
  rgb(3pt)=(0.02275049,0.02554464,0.11108639);
  rgb(4pt)=(0.02759119,0.02818316,0.11483751);
  rgb(5pt)=(0.03285175,0.03088792,0.11863035);
  rgb(6pt)=(0.03853466,0.03365771,0.12245873);
  rgb(7pt)=(0.04447016,0.03648425,0.12631831);
  rgb(8pt)=(0.05032105,0.03936808,0.13020508);
  rgb(9pt)=(0.05611171,0.04224835,0.13411624);
  rgb(10pt)=(0.0618531,0.04504866,0.13804929);
  rgb(11pt)=(0.06755457,0.04778179,0.14200206);
  rgb(12pt)=(0.0732236,0.05045047,0.14597263);
  rgb(13pt)=(0.0788708,0.05305461,0.14995981);
  rgb(14pt)=(0.08450105,0.05559631,0.15396203);
  rgb(15pt)=(0.09011319,0.05808059,0.15797687);
  rgb(16pt)=(0.09572396,0.06050127,0.16200507);
  rgb(17pt)=(0.10132312,0.06286782,0.16604287);
  rgb(18pt)=(0.10692823,0.06517224,0.17009175);
  rgb(19pt)=(0.1125315,0.06742194,0.17414848);
  rgb(20pt)=(0.11813947,0.06961499,0.17821272);
  rgb(21pt)=(0.12375803,0.07174938,0.18228425);
  rgb(22pt)=(0.12938228,0.07383015,0.18636053);
  rgb(23pt)=(0.13501631,0.07585609,0.19044109);
  rgb(24pt)=(0.14066867,0.0778224,0.19452676);
  rgb(25pt)=(0.14633406,0.07973393,0.1986151);
  rgb(26pt)=(0.15201338,0.08159108,0.20270523);
  rgb(27pt)=(0.15770877,0.08339312,0.20679668);
  rgb(28pt)=(0.16342174,0.0851396,0.21088893);
  rgb(29pt)=(0.16915387,0.08682996,0.21498104);
  rgb(30pt)=(0.17489524,0.08848235,0.2190294);
  rgb(31pt)=(0.18065495,0.09009031,0.22303512);
  rgb(32pt)=(0.18643324,0.09165431,0.22699705);
  rgb(33pt)=(0.19223028,0.09317479,0.23091409);
  rgb(34pt)=(0.19804623,0.09465217,0.23478512);
  rgb(35pt)=(0.20388117,0.09608689,0.23860907);
  rgb(36pt)=(0.20973515,0.09747934,0.24238489);
  rgb(37pt)=(0.21560818,0.09882993,0.24611154);
  rgb(38pt)=(0.22150014,0.10013944,0.2497868);
  rgb(39pt)=(0.22741085,0.10140876,0.25340813);
  rgb(40pt)=(0.23334047,0.10263737,0.25697736);
  rgb(41pt)=(0.23928891,0.10382562,0.2604936);
  rgb(42pt)=(0.24525608,0.10497384,0.26395596);
  rgb(43pt)=(0.25124182,0.10608236,0.26736359);
  rgb(44pt)=(0.25724602,0.10715148,0.27071569);
  rgb(45pt)=(0.26326851,0.1081815,0.27401148);
  rgb(46pt)=(0.26930915,0.1091727,0.2772502);
  rgb(47pt)=(0.27536766,0.11012568,0.28043021);
  rgb(48pt)=(0.28144375,0.11104133,0.2835489);
  rgb(49pt)=(0.2875374,0.11191896,0.28660853);
  rgb(50pt)=(0.29364846,0.11275876,0.2896085);
  rgb(51pt)=(0.29977678,0.11356089,0.29254823);
  rgb(52pt)=(0.30592213,0.11432553,0.29542718);
  rgb(53pt)=(0.31208435,0.11505284,0.29824485);
  rgb(54pt)=(0.31826327,0.1157429,0.30100076);
  rgb(55pt)=(0.32445869,0.11639585,0.30369448);
  rgb(56pt)=(0.33067031,0.11701189,0.30632563);
  rgb(57pt)=(0.33689808,0.11759095,0.3088938);
  rgb(58pt)=(0.34314168,0.11813362,0.31139721);
  rgb(59pt)=(0.34940101,0.11863987,0.3138355);
  rgb(60pt)=(0.355676,0.11910909,0.31620996);
  rgb(61pt)=(0.36196644,0.1195413,0.31852037);
  rgb(62pt)=(0.36827206,0.11993653,0.32076656);
  rgb(63pt)=(0.37459292,0.12029443,0.32294825);
  rgb(64pt)=(0.38092887,0.12061482,0.32506528);
  rgb(65pt)=(0.38727975,0.12089756,0.3271175);
  rgb(66pt)=(0.39364518,0.12114272,0.32910494);
  rgb(67pt)=(0.40002537,0.12134964,0.33102734);
  rgb(68pt)=(0.40642019,0.12151801,0.33288464);
  rgb(69pt)=(0.41282936,0.12164769,0.33467689);
  rgb(70pt)=(0.41925278,0.12173833,0.33640407);
  rgb(71pt)=(0.42569057,0.12178916,0.33806605);
  rgb(72pt)=(0.43214263,0.12179973,0.33966284);
  rgb(73pt)=(0.43860848,0.12177004,0.34119475);
  rgb(74pt)=(0.44508855,0.12169883,0.34266151);
  rgb(75pt)=(0.45158266,0.12158557,0.34406324);
  rgb(76pt)=(0.45809049,0.12142996,0.34540024);
  rgb(77pt)=(0.46461238,0.12123063,0.34667231);
  rgb(78pt)=(0.47114798,0.12098721,0.34787978);
  rgb(79pt)=(0.47769736,0.12069864,0.34902273);
  rgb(80pt)=(0.48426077,0.12036349,0.35010104);
  rgb(81pt)=(0.49083761,0.11998161,0.35111537);
  rgb(82pt)=(0.49742847,0.11955087,0.35206533);
  rgb(83pt)=(0.50403286,0.11907081,0.35295152);
  rgb(84pt)=(0.51065109,0.11853959,0.35377385);
  rgb(85pt)=(0.51728314,0.1179558,0.35453252);
  rgb(86pt)=(0.52392883,0.11731817,0.35522789);
  rgb(87pt)=(0.53058853,0.11662445,0.35585982);
  rgb(88pt)=(0.53726173,0.11587369,0.35642903);
  rgb(89pt)=(0.54394898,0.11506307,0.35693521);
  rgb(90pt)=(0.5506426,0.11420757,0.35737863);
  rgb(91pt)=(0.55734473,0.11330456,0.35775059);
  rgb(92pt)=(0.56405586,0.11235265,0.35804813);
  rgb(93pt)=(0.57077365,0.11135597,0.35827146);
  rgb(94pt)=(0.5774991,0.11031233,0.35841679);
  rgb(95pt)=(0.58422945,0.10922707,0.35848469);
  rgb(96pt)=(0.59096382,0.10810205,0.35847347);
  rgb(97pt)=(0.59770215,0.10693774,0.35838029);
  rgb(98pt)=(0.60444226,0.10573912,0.35820487);
  rgb(99pt)=(0.61118304,0.10450943,0.35794557);
  rgb(100pt)=(0.61792306,0.10325288,0.35760108);
  rgb(101pt)=(0.62466162,0.10197244,0.35716891);
  rgb(102pt)=(0.63139686,0.10067417,0.35664819);
  rgb(103pt)=(0.63812122,0.09938212,0.35603757);
  rgb(104pt)=(0.64483795,0.0980891,0.35533555);
  rgb(105pt)=(0.65154562,0.09680192,0.35454107);
  rgb(106pt)=(0.65824241,0.09552918,0.3536529);
  rgb(107pt)=(0.66492652,0.09428017,0.3526697);
  rgb(108pt)=(0.67159578,0.09306598,0.35159077);
  rgb(109pt)=(0.67824099,0.09192342,0.3504148);
  rgb(110pt)=(0.684863,0.09085633,0.34914061);
  rgb(111pt)=(0.69146268,0.0898675,0.34776864);
  rgb(112pt)=(0.69803757,0.08897226,0.3462986);
  rgb(113pt)=(0.70457834,0.0882129,0.34473046);
  rgb(114pt)=(0.71108138,0.08761223,0.3430635);
  rgb(115pt)=(0.7175507,0.08716212,0.34129974);
  rgb(116pt)=(0.72398193,0.08688725,0.33943958);
  rgb(117pt)=(0.73035829,0.0868623,0.33748452);
  rgb(118pt)=(0.73669146,0.08704683,0.33543669);
  rgb(119pt)=(0.74297501,0.08747196,0.33329799);
  rgb(120pt)=(0.74919318,0.08820542,0.33107204);
  rgb(121pt)=(0.75535825,0.08919792,0.32876184);
  rgb(122pt)=(0.76145589,0.09050716,0.32637117);
  rgb(123pt)=(0.76748424,0.09213602,0.32390525);
  rgb(124pt)=(0.77344838,0.09405684,0.32136808);
  rgb(125pt)=(0.77932641,0.09634794,0.31876642);
  rgb(126pt)=(0.78513609,0.09892473,0.31610488);
  rgb(127pt)=(0.79085854,0.10184672,0.313391);
  rgb(128pt)=(0.7965014,0.10506637,0.31063031);
  rgb(129pt)=(0.80205987,0.10858333,0.30783);
  rgb(130pt)=(0.80752799,0.11239964,0.30499738);
  rgb(131pt)=(0.81291606,0.11645784,0.30213802);
  rgb(132pt)=(0.81820481,0.12080606,0.29926105);
  rgb(133pt)=(0.82341472,0.12535343,0.2963705);
  rgb(134pt)=(0.82852822,0.13014118,0.29347474);
  rgb(135pt)=(0.83355779,0.13511035,0.29057852);
  rgb(136pt)=(0.83850183,0.14025098,0.2876878);
  rgb(137pt)=(0.84335441,0.14556683,0.28480819);
  rgb(138pt)=(0.84813096,0.15099892,0.281943);
  rgb(139pt)=(0.85281737,0.15657772,0.27909826);
  rgb(140pt)=(0.85742602,0.1622583,0.27627462);
  rgb(141pt)=(0.86196552,0.16801239,0.27346473);
  rgb(142pt)=(0.86641628,0.17387796,0.27070818);
  rgb(143pt)=(0.87079129,0.17982114,0.26797378);
  rgb(144pt)=(0.87507281,0.18587368,0.26529697);
  rgb(145pt)=(0.87925878,0.19203259,0.26268136);
  rgb(146pt)=(0.8833417,0.19830556,0.26014181);
  rgb(147pt)=(0.88731387,0.20469941,0.25769539);
  rgb(148pt)=(0.89116859,0.21121788,0.2553592);
  rgb(149pt)=(0.89490337,0.21785614,0.25314362);
  rgb(150pt)=(0.8985026,0.22463251,0.25108745);
  rgb(151pt)=(0.90197527,0.23152063,0.24918223);
  rgb(152pt)=(0.90530097,0.23854541,0.24748098);
  rgb(153pt)=(0.90848638,0.24568473,0.24598324);
  rgb(154pt)=(0.911533,0.25292623,0.24470258);
  rgb(155pt)=(0.9144225,0.26028902,0.24369359);
  rgb(156pt)=(0.91717106,0.26773821,0.24294137);
  rgb(157pt)=(0.91978131,0.27526191,0.24245973);
  rgb(158pt)=(0.92223947,0.28287251,0.24229568);
  rgb(159pt)=(0.92456587,0.29053388,0.24242622);
  rgb(160pt)=(0.92676657,0.29823282,0.24285536);
  rgb(161pt)=(0.92882964,0.30598085,0.24362274);
  rgb(162pt)=(0.93078135,0.31373977,0.24468803);
  rgb(163pt)=(0.93262051,0.3215093,0.24606461);
  rgb(164pt)=(0.93435067,0.32928362,0.24775328);
  rgb(165pt)=(0.93599076,0.33703942,0.24972157);
  rgb(166pt)=(0.93752831,0.34479177,0.25199928);
  rgb(167pt)=(0.93899289,0.35250734,0.25452808);
  rgb(168pt)=(0.94036561,0.36020899,0.25734661);
  rgb(169pt)=(0.94167588,0.36786594,0.2603949);
  rgb(170pt)=(0.94291042,0.37549479,0.26369821);
  rgb(171pt)=(0.94408513,0.3830811,0.26722004);
  rgb(172pt)=(0.94520419,0.39062329,0.27094924);
  rgb(173pt)=(0.94625977,0.39813168,0.27489742);
  rgb(174pt)=(0.94727016,0.4055909,0.27902322);
  rgb(175pt)=(0.94823505,0.41300424,0.28332283);
  rgb(176pt)=(0.94914549,0.42038251,0.28780969);
  rgb(177pt)=(0.95001704,0.42771398,0.29244728);
  rgb(178pt)=(0.95085121,0.43500005,0.29722817);
  rgb(179pt)=(0.95165009,0.44224144,0.30214494);
  rgb(180pt)=(0.9524044,0.44944853,0.3072105);
  rgb(181pt)=(0.95312556,0.45661389,0.31239776);
  rgb(182pt)=(0.95381595,0.46373781,0.31769923);
  rgb(183pt)=(0.95447591,0.47082238,0.32310953);
  rgb(184pt)=(0.95510255,0.47787236,0.32862553);
  rgb(185pt)=(0.95569679,0.48489115,0.33421404);
  rgb(186pt)=(0.95626788,0.49187351,0.33985601);
  rgb(187pt)=(0.95681685,0.49882008,0.34555431);
  rgb(188pt)=(0.9573439,0.50573243,0.35130912);
  rgb(189pt)=(0.95784842,0.51261283,0.35711942);
  rgb(190pt)=(0.95833051,0.51946267,0.36298589);
  rgb(191pt)=(0.95879054,0.52628305,0.36890904);
  rgb(192pt)=(0.95922872,0.53307513,0.3748895);
  rgb(193pt)=(0.95964538,0.53983991,0.38092784);
  rgb(194pt)=(0.96004345,0.54657593,0.3870292);
  rgb(195pt)=(0.96042097,0.55328624,0.39319057);
  rgb(196pt)=(0.96077819,0.55997184,0.39941173);
  rgb(197pt)=(0.9611152,0.5666337,0.40569343);
  rgb(198pt)=(0.96143273,0.57327231,0.41203603);
  rgb(199pt)=(0.96173392,0.57988594,0.41844491);
  rgb(200pt)=(0.96201757,0.58647675,0.42491751);
  rgb(201pt)=(0.96228344,0.59304598,0.43145271);
  rgb(202pt)=(0.96253168,0.5995944,0.43805131);
  rgb(203pt)=(0.96276513,0.60612062,0.44471698);
  rgb(204pt)=(0.96298491,0.6126247,0.45145074);
  rgb(205pt)=(0.96318967,0.61910879,0.45824902);
  rgb(206pt)=(0.96337949,0.6255736,0.46511271);
  rgb(207pt)=(0.96355923,0.63201624,0.47204746);
  rgb(208pt)=(0.96372785,0.63843852,0.47905028);
  rgb(209pt)=(0.96388426,0.64484214,0.4861196);
  rgb(210pt)=(0.96403203,0.65122535,0.4932578);
  rgb(211pt)=(0.96417332,0.65758729,0.50046894);
  rgb(212pt)=(0.9643063,0.66393045,0.5077467);
  rgb(213pt)=(0.96443322,0.67025402,0.51509334);
  rgb(214pt)=(0.96455845,0.67655564,0.52251447);
  rgb(215pt)=(0.96467922,0.68283846,0.53000231);
  rgb(216pt)=(0.96479861,0.68910113,0.53756026);
  rgb(217pt)=(0.96492035,0.69534192,0.5451917);
  rgb(218pt)=(0.96504223,0.7015636,0.5528892);
  rgb(219pt)=(0.96516917,0.70776351,0.5606593);
  rgb(220pt)=(0.96530224,0.71394212,0.56849894);
  rgb(221pt)=(0.96544032,0.72010124,0.57640375);
  rgb(222pt)=(0.96559206,0.72623592,0.58438387);
  rgb(223pt)=(0.96575293,0.73235058,0.59242739);
  rgb(224pt)=(0.96592829,0.73844258,0.60053991);
  rgb(225pt)=(0.96612013,0.74451182,0.60871954);
  rgb(226pt)=(0.96632832,0.75055966,0.61696136);
  rgb(227pt)=(0.96656022,0.75658231,0.62527295);
  rgb(228pt)=(0.96681185,0.76258381,0.63364277);
  rgb(229pt)=(0.96709183,0.76855969,0.64207921);
  rgb(230pt)=(0.96739773,0.77451297,0.65057302);
  rgb(231pt)=(0.96773482,0.78044149,0.65912731);
  rgb(232pt)=(0.96810471,0.78634563,0.66773889);
  rgb(233pt)=(0.96850919,0.79222565,0.6764046);
  rgb(234pt)=(0.96893132,0.79809112,0.68512266);
  rgb(235pt)=(0.96935926,0.80395415,0.69383201);
  rgb(236pt)=(0.9698028,0.80981139,0.70252255);
  rgb(237pt)=(0.97025511,0.81566605,0.71120296);
  rgb(238pt)=(0.97071849,0.82151775,0.71987163);
  rgb(239pt)=(0.97120159,0.82736371,0.72851999);
  rgb(240pt)=(0.97169389,0.83320847,0.73716071);
  rgb(241pt)=(0.97220061,0.83905052,0.74578903);
  rgb(242pt)=(0.97272597,0.84488881,0.75440141);
  rgb(243pt)=(0.97327085,0.85072354,0.76299805);
  rgb(244pt)=(0.97383206,0.85655639,0.77158353);
  rgb(245pt)=(0.97441222,0.86238689,0.78015619);
  rgb(246pt)=(0.97501782,0.86821321,0.78871034);
  rgb(247pt)=(0.97564391,0.87403763,0.79725261);
  rgb(248pt)=(0.97628674,0.87986189,0.8057883);
  rgb(249pt)=(0.97696114,0.88568129,0.81430324);
  rgb(250pt)=(0.97765722,0.89149971,0.82280948);
  rgb(251pt)=(0.97837585,0.89731727,0.83130786);
  rgb(252pt)=(0.97912374,0.90313207,0.83979337);
  rgb(253pt)=(0.979891,0.90894778,0.84827858);
  rgb(254pt)=(0.98067764,0.91476465,0.85676611);
  rgb(255pt)=(0.98137749,0.92061729,0.86536915)
},
point meta max=3.8101671442299,
point meta min=-2.41594987562462,
tick align=outside,
tick pos=left,
x grid style={darkgray176},
xlabel={$\theta_1$},
xmin=0, xmax=40,
xtick style={color=black},
xtick={0,20,40},
xticklabels={$0$,$\pi$,$2\pi$},
y dir=reverse,
y grid style={darkgray176},
ylabel={$\theta_0$},
ymin=0, ymax=40,
ytick style={color=black},
ytick={0,20,40},
yticklabel style={rotate=90.0},
yticklabels={$2\pi$,$\pi$,$0$}
]
\addplot graphics[xmin=0, xmax=40, ymin=40, ymax=0] {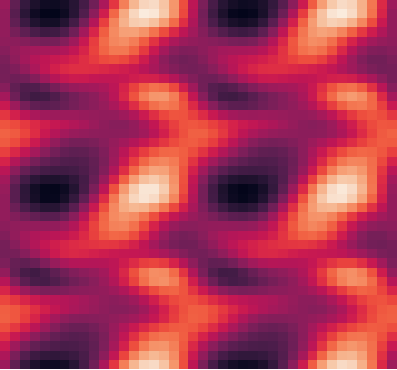};
\end{axis}

\end{tikzpicture}

%% file: tikz_figures/hobo_qaoa_landscape.tex
% This file was created with tikzplotlib v0.10.1.
\begin{tikzpicture}

\definecolor{darkgray176}{RGB}{176,176,176}

\begin{axis}[
colorbar,
colorbar style={ylabel={Energy}, ylabel style={yshift=-0.1cm}},
colormap={mymap}{[1pt]
 rgb(0pt)=(0.01060815,0.01808215,0.10018654);
  rgb(1pt)=(0.01428972,0.02048237,0.10374486);
  rgb(2pt)=(0.01831941,0.0229766,0.10738511);
  rgb(3pt)=(0.02275049,0.02554464,0.11108639);
  rgb(4pt)=(0.02759119,0.02818316,0.11483751);
  rgb(5pt)=(0.03285175,0.03088792,0.11863035);
  rgb(6pt)=(0.03853466,0.03365771,0.12245873);
  rgb(7pt)=(0.04447016,0.03648425,0.12631831);
  rgb(8pt)=(0.05032105,0.03936808,0.13020508);
  rgb(9pt)=(0.05611171,0.04224835,0.13411624);
  rgb(10pt)=(0.0618531,0.04504866,0.13804929);
  rgb(11pt)=(0.06755457,0.04778179,0.14200206);
  rgb(12pt)=(0.0732236,0.05045047,0.14597263);
  rgb(13pt)=(0.0788708,0.05305461,0.14995981);
  rgb(14pt)=(0.08450105,0.05559631,0.15396203);
  rgb(15pt)=(0.09011319,0.05808059,0.15797687);
  rgb(16pt)=(0.09572396,0.06050127,0.16200507);
  rgb(17pt)=(0.10132312,0.06286782,0.16604287);
  rgb(18pt)=(0.10692823,0.06517224,0.17009175);
  rgb(19pt)=(0.1125315,0.06742194,0.17414848);
  rgb(20pt)=(0.11813947,0.06961499,0.17821272);
  rgb(21pt)=(0.12375803,0.07174938,0.18228425);
  rgb(22pt)=(0.12938228,0.07383015,0.18636053);
  rgb(23pt)=(0.13501631,0.07585609,0.19044109);
  rgb(24pt)=(0.14066867,0.0778224,0.19452676);
  rgb(25pt)=(0.14633406,0.07973393,0.1986151);
  rgb(26pt)=(0.15201338,0.08159108,0.20270523);
  rgb(27pt)=(0.15770877,0.08339312,0.20679668);
  rgb(28pt)=(0.16342174,0.0851396,0.21088893);
  rgb(29pt)=(0.16915387,0.08682996,0.21498104);
  rgb(30pt)=(0.17489524,0.08848235,0.2190294);
  rgb(31pt)=(0.18065495,0.09009031,0.22303512);
  rgb(32pt)=(0.18643324,0.09165431,0.22699705);
  rgb(33pt)=(0.19223028,0.09317479,0.23091409);
  rgb(34pt)=(0.19804623,0.09465217,0.23478512);
  rgb(35pt)=(0.20388117,0.09608689,0.23860907);
  rgb(36pt)=(0.20973515,0.09747934,0.24238489);
  rgb(37pt)=(0.21560818,0.09882993,0.24611154);
  rgb(38pt)=(0.22150014,0.10013944,0.2497868);
  rgb(39pt)=(0.22741085,0.10140876,0.25340813);
  rgb(40pt)=(0.23334047,0.10263737,0.25697736);
  rgb(41pt)=(0.23928891,0.10382562,0.2604936);
  rgb(42pt)=(0.24525608,0.10497384,0.26395596);
  rgb(43pt)=(0.25124182,0.10608236,0.26736359);
  rgb(44pt)=(0.25724602,0.10715148,0.27071569);
  rgb(45pt)=(0.26326851,0.1081815,0.27401148);
  rgb(46pt)=(0.26930915,0.1091727,0.2772502);
  rgb(47pt)=(0.27536766,0.11012568,0.28043021);
  rgb(48pt)=(0.28144375,0.11104133,0.2835489);
  rgb(49pt)=(0.2875374,0.11191896,0.28660853);
  rgb(50pt)=(0.29364846,0.11275876,0.2896085);
  rgb(51pt)=(0.29977678,0.11356089,0.29254823);
  rgb(52pt)=(0.30592213,0.11432553,0.29542718);
  rgb(53pt)=(0.31208435,0.11505284,0.29824485);
  rgb(54pt)=(0.31826327,0.1157429,0.30100076);
  rgb(55pt)=(0.32445869,0.11639585,0.30369448);
  rgb(56pt)=(0.33067031,0.11701189,0.30632563);
  rgb(57pt)=(0.33689808,0.11759095,0.3088938);
  rgb(58pt)=(0.34314168,0.11813362,0.31139721);
  rgb(59pt)=(0.34940101,0.11863987,0.3138355);
  rgb(60pt)=(0.355676,0.11910909,0.31620996);
  rgb(61pt)=(0.36196644,0.1195413,0.31852037);
  rgb(62pt)=(0.36827206,0.11993653,0.32076656);
  rgb(63pt)=(0.37459292,0.12029443,0.32294825);
  rgb(64pt)=(0.38092887,0.12061482,0.32506528);
  rgb(65pt)=(0.38727975,0.12089756,0.3271175);
  rgb(66pt)=(0.39364518,0.12114272,0.32910494);
  rgb(67pt)=(0.40002537,0.12134964,0.33102734);
  rgb(68pt)=(0.40642019,0.12151801,0.33288464);
  rgb(69pt)=(0.41282936,0.12164769,0.33467689);
  rgb(70pt)=(0.41925278,0.12173833,0.33640407);
  rgb(71pt)=(0.42569057,0.12178916,0.33806605);
  rgb(72pt)=(0.43214263,0.12179973,0.33966284);
  rgb(73pt)=(0.43860848,0.12177004,0.34119475);
  rgb(74pt)=(0.44508855,0.12169883,0.34266151);
  rgb(75pt)=(0.45158266,0.12158557,0.34406324);
  rgb(76pt)=(0.45809049,0.12142996,0.34540024);
  rgb(77pt)=(0.46461238,0.12123063,0.34667231);
  rgb(78pt)=(0.47114798,0.12098721,0.34787978);
  rgb(79pt)=(0.47769736,0.12069864,0.34902273);
  rgb(80pt)=(0.48426077,0.12036349,0.35010104);
  rgb(81pt)=(0.49083761,0.11998161,0.35111537);
  rgb(82pt)=(0.49742847,0.11955087,0.35206533);
  rgb(83pt)=(0.50403286,0.11907081,0.35295152);
  rgb(84pt)=(0.51065109,0.11853959,0.35377385);
  rgb(85pt)=(0.51728314,0.1179558,0.35453252);
  rgb(86pt)=(0.52392883,0.11731817,0.35522789);
  rgb(87pt)=(0.53058853,0.11662445,0.35585982);
  rgb(88pt)=(0.53726173,0.11587369,0.35642903);
  rgb(89pt)=(0.54394898,0.11506307,0.35693521);
  rgb(90pt)=(0.5506426,0.11420757,0.35737863);
  rgb(91pt)=(0.55734473,0.11330456,0.35775059);
  rgb(92pt)=(0.56405586,0.11235265,0.35804813);
  rgb(93pt)=(0.57077365,0.11135597,0.35827146);
  rgb(94pt)=(0.5774991,0.11031233,0.35841679);
  rgb(95pt)=(0.58422945,0.10922707,0.35848469);
  rgb(96pt)=(0.59096382,0.10810205,0.35847347);
  rgb(97pt)=(0.59770215,0.10693774,0.35838029);
  rgb(98pt)=(0.60444226,0.10573912,0.35820487);
  rgb(99pt)=(0.61118304,0.10450943,0.35794557);
  rgb(100pt)=(0.61792306,0.10325288,0.35760108);
  rgb(101pt)=(0.62466162,0.10197244,0.35716891);
  rgb(102pt)=(0.63139686,0.10067417,0.35664819);
  rgb(103pt)=(0.63812122,0.09938212,0.35603757);
  rgb(104pt)=(0.64483795,0.0980891,0.35533555);
  rgb(105pt)=(0.65154562,0.09680192,0.35454107);
  rgb(106pt)=(0.65824241,0.09552918,0.3536529);
  rgb(107pt)=(0.66492652,0.09428017,0.3526697);
  rgb(108pt)=(0.67159578,0.09306598,0.35159077);
  rgb(109pt)=(0.67824099,0.09192342,0.3504148);
  rgb(110pt)=(0.684863,0.09085633,0.34914061);
  rgb(111pt)=(0.69146268,0.0898675,0.34776864);
  rgb(112pt)=(0.69803757,0.08897226,0.3462986);
  rgb(113pt)=(0.70457834,0.0882129,0.34473046);
  rgb(114pt)=(0.71108138,0.08761223,0.3430635);
  rgb(115pt)=(0.7175507,0.08716212,0.34129974);
  rgb(116pt)=(0.72398193,0.08688725,0.33943958);
  rgb(117pt)=(0.73035829,0.0868623,0.33748452);
  rgb(118pt)=(0.73669146,0.08704683,0.33543669);
  rgb(119pt)=(0.74297501,0.08747196,0.33329799);
  rgb(120pt)=(0.74919318,0.08820542,0.33107204);
  rgb(121pt)=(0.75535825,0.08919792,0.32876184);
  rgb(122pt)=(0.76145589,0.09050716,0.32637117);
  rgb(123pt)=(0.76748424,0.09213602,0.32390525);
  rgb(124pt)=(0.77344838,0.09405684,0.32136808);
  rgb(125pt)=(0.77932641,0.09634794,0.31876642);
  rgb(126pt)=(0.78513609,0.09892473,0.31610488);
  rgb(127pt)=(0.79085854,0.10184672,0.313391);
  rgb(128pt)=(0.7965014,0.10506637,0.31063031);
  rgb(129pt)=(0.80205987,0.10858333,0.30783);
  rgb(130pt)=(0.80752799,0.11239964,0.30499738);
  rgb(131pt)=(0.81291606,0.11645784,0.30213802);
  rgb(132pt)=(0.81820481,0.12080606,0.29926105);
  rgb(133pt)=(0.82341472,0.12535343,0.2963705);
  rgb(134pt)=(0.82852822,0.13014118,0.29347474);
  rgb(135pt)=(0.83355779,0.13511035,0.29057852);
  rgb(136pt)=(0.83850183,0.14025098,0.2876878);
  rgb(137pt)=(0.84335441,0.14556683,0.28480819);
  rgb(138pt)=(0.84813096,0.15099892,0.281943);
  rgb(139pt)=(0.85281737,0.15657772,0.27909826);
  rgb(140pt)=(0.85742602,0.1622583,0.27627462);
  rgb(141pt)=(0.86196552,0.16801239,0.27346473);
  rgb(142pt)=(0.86641628,0.17387796,0.27070818);
  rgb(143pt)=(0.87079129,0.17982114,0.26797378);
  rgb(144pt)=(0.87507281,0.18587368,0.26529697);
  rgb(145pt)=(0.87925878,0.19203259,0.26268136);
  rgb(146pt)=(0.8833417,0.19830556,0.26014181);
  rgb(147pt)=(0.88731387,0.20469941,0.25769539);
  rgb(148pt)=(0.89116859,0.21121788,0.2553592);
  rgb(149pt)=(0.89490337,0.21785614,0.25314362);
  rgb(150pt)=(0.8985026,0.22463251,0.25108745);
  rgb(151pt)=(0.90197527,0.23152063,0.24918223);
  rgb(152pt)=(0.90530097,0.23854541,0.24748098);
  rgb(153pt)=(0.90848638,0.24568473,0.24598324);
  rgb(154pt)=(0.911533,0.25292623,0.24470258);
  rgb(155pt)=(0.9144225,0.26028902,0.24369359);
  rgb(156pt)=(0.91717106,0.26773821,0.24294137);
  rgb(157pt)=(0.91978131,0.27526191,0.24245973);
  rgb(158pt)=(0.92223947,0.28287251,0.24229568);
  rgb(159pt)=(0.92456587,0.29053388,0.24242622);
  rgb(160pt)=(0.92676657,0.29823282,0.24285536);
  rgb(161pt)=(0.92882964,0.30598085,0.24362274);
  rgb(162pt)=(0.93078135,0.31373977,0.24468803);
  rgb(163pt)=(0.93262051,0.3215093,0.24606461);
  rgb(164pt)=(0.93435067,0.32928362,0.24775328);
  rgb(165pt)=(0.93599076,0.33703942,0.24972157);
  rgb(166pt)=(0.93752831,0.34479177,0.25199928);
  rgb(167pt)=(0.93899289,0.35250734,0.25452808);
  rgb(168pt)=(0.94036561,0.36020899,0.25734661);
  rgb(169pt)=(0.94167588,0.36786594,0.2603949);
  rgb(170pt)=(0.94291042,0.37549479,0.26369821);
  rgb(171pt)=(0.94408513,0.3830811,0.26722004);
  rgb(172pt)=(0.94520419,0.39062329,0.27094924);
  rgb(173pt)=(0.94625977,0.39813168,0.27489742);
  rgb(174pt)=(0.94727016,0.4055909,0.27902322);
  rgb(175pt)=(0.94823505,0.41300424,0.28332283);
  rgb(176pt)=(0.94914549,0.42038251,0.28780969);
  rgb(177pt)=(0.95001704,0.42771398,0.29244728);
  rgb(178pt)=(0.95085121,0.43500005,0.29722817);
  rgb(179pt)=(0.95165009,0.44224144,0.30214494);
  rgb(180pt)=(0.9524044,0.44944853,0.3072105);
  rgb(181pt)=(0.95312556,0.45661389,0.31239776);
  rgb(182pt)=(0.95381595,0.46373781,0.31769923);
  rgb(183pt)=(0.95447591,0.47082238,0.32310953);
  rgb(184pt)=(0.95510255,0.47787236,0.32862553);
  rgb(185pt)=(0.95569679,0.48489115,0.33421404);
  rgb(186pt)=(0.95626788,0.49187351,0.33985601);
  rgb(187pt)=(0.95681685,0.49882008,0.34555431);
  rgb(188pt)=(0.9573439,0.50573243,0.35130912);
  rgb(189pt)=(0.95784842,0.51261283,0.35711942);
  rgb(190pt)=(0.95833051,0.51946267,0.36298589);
  rgb(191pt)=(0.95879054,0.52628305,0.36890904);
  rgb(192pt)=(0.95922872,0.53307513,0.3748895);
  rgb(193pt)=(0.95964538,0.53983991,0.38092784);
  rgb(194pt)=(0.96004345,0.54657593,0.3870292);
  rgb(195pt)=(0.96042097,0.55328624,0.39319057);
  rgb(196pt)=(0.96077819,0.55997184,0.39941173);
  rgb(197pt)=(0.9611152,0.5666337,0.40569343);
  rgb(198pt)=(0.96143273,0.57327231,0.41203603);
  rgb(199pt)=(0.96173392,0.57988594,0.41844491);
  rgb(200pt)=(0.96201757,0.58647675,0.42491751);
  rgb(201pt)=(0.96228344,0.59304598,0.43145271);
  rgb(202pt)=(0.96253168,0.5995944,0.43805131);
  rgb(203pt)=(0.96276513,0.60612062,0.44471698);
  rgb(204pt)=(0.96298491,0.6126247,0.45145074);
  rgb(205pt)=(0.96318967,0.61910879,0.45824902);
  rgb(206pt)=(0.96337949,0.6255736,0.46511271);
  rgb(207pt)=(0.96355923,0.63201624,0.47204746);
  rgb(208pt)=(0.96372785,0.63843852,0.47905028);
  rgb(209pt)=(0.96388426,0.64484214,0.4861196);
  rgb(210pt)=(0.96403203,0.65122535,0.4932578);
  rgb(211pt)=(0.96417332,0.65758729,0.50046894);
  rgb(212pt)=(0.9643063,0.66393045,0.5077467);
  rgb(213pt)=(0.96443322,0.67025402,0.51509334);
  rgb(214pt)=(0.96455845,0.67655564,0.52251447);
  rgb(215pt)=(0.96467922,0.68283846,0.53000231);
  rgb(216pt)=(0.96479861,0.68910113,0.53756026);
  rgb(217pt)=(0.96492035,0.69534192,0.5451917);
  rgb(218pt)=(0.96504223,0.7015636,0.5528892);
  rgb(219pt)=(0.96516917,0.70776351,0.5606593);
  rgb(220pt)=(0.96530224,0.71394212,0.56849894);
  rgb(221pt)=(0.96544032,0.72010124,0.57640375);
  rgb(222pt)=(0.96559206,0.72623592,0.58438387);
  rgb(223pt)=(0.96575293,0.73235058,0.59242739);
  rgb(224pt)=(0.96592829,0.73844258,0.60053991);
  rgb(225pt)=(0.96612013,0.74451182,0.60871954);
  rgb(226pt)=(0.96632832,0.75055966,0.61696136);
  rgb(227pt)=(0.96656022,0.75658231,0.62527295);
  rgb(228pt)=(0.96681185,0.76258381,0.63364277);
  rgb(229pt)=(0.96709183,0.76855969,0.64207921);
  rgb(230pt)=(0.96739773,0.77451297,0.65057302);
  rgb(231pt)=(0.96773482,0.78044149,0.65912731);
  rgb(232pt)=(0.96810471,0.78634563,0.66773889);
  rgb(233pt)=(0.96850919,0.79222565,0.6764046);
  rgb(234pt)=(0.96893132,0.79809112,0.68512266);
  rgb(235pt)=(0.96935926,0.80395415,0.69383201);
  rgb(236pt)=(0.9698028,0.80981139,0.70252255);
  rgb(237pt)=(0.97025511,0.81566605,0.71120296);
  rgb(238pt)=(0.97071849,0.82151775,0.71987163);
  rgb(239pt)=(0.97120159,0.82736371,0.72851999);
  rgb(240pt)=(0.97169389,0.83320847,0.73716071);
  rgb(241pt)=(0.97220061,0.83905052,0.74578903);
  rgb(242pt)=(0.97272597,0.84488881,0.75440141);
  rgb(243pt)=(0.97327085,0.85072354,0.76299805);
  rgb(244pt)=(0.97383206,0.85655639,0.77158353);
  rgb(245pt)=(0.97441222,0.86238689,0.78015619);
  rgb(246pt)=(0.97501782,0.86821321,0.78871034);
  rgb(247pt)=(0.97564391,0.87403763,0.79725261);
  rgb(248pt)=(0.97628674,0.87986189,0.8057883);
  rgb(249pt)=(0.97696114,0.88568129,0.81430324);
  rgb(250pt)=(0.97765722,0.89149971,0.82280948);
  rgb(251pt)=(0.97837585,0.89731727,0.83130786);
  rgb(252pt)=(0.97912374,0.90313207,0.83979337);
  rgb(253pt)=(0.979891,0.90894778,0.84827858);
  rgb(254pt)=(0.98067764,0.91476465,0.85676611);
  rgb(255pt)=(0.98137749,0.92061729,0.86536915)
},
point meta max=2.92271491901595,
point meta min=2.30937321928004,
tick align=outside,
tick pos=left,
x grid style={darkgray176},
xlabel={$\theta_1$},
xmin=0, xmax=40,
xtick style={color=black},
xtick={0,20,40},
xticklabels={$0$,$\pi$,$2\pi$},
y dir=reverse,
y grid style={darkgray176},
ylabel={$\theta_0$},
ymin=0, ymax=40,
ytick style={color=black},
ytick={0,20,40},
yticklabel style={rotate=90.0},
yticklabels={$2\pi$,$\pi$,$0$}
]
\addplot graphics[xmin=0, xmax=40, ymin=40, ymax=0] {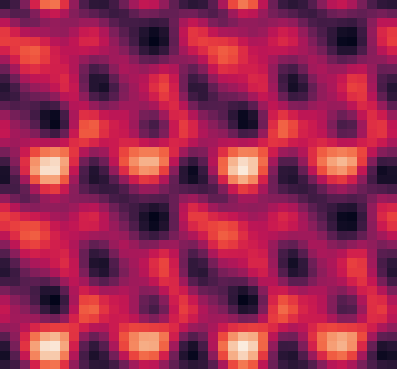};
\end{axis}

\end{tikzpicture}